\documentclass[a4paper,11pt]{article}

\usepackage{graphicx}

\usepackage[all]{xy}
\usepackage{latexsym,amssymb,amsmath,amsfonts, amsthm, xcolor}

\usepackage[all]{xy}
\usepackage{latexsym,amssymb,amsmath,amsfonts, amsthm, xcolor}

\def\vbar{\mathchoice{\vrule height6.3ptdepth-.5ptwidth.8pt\kern-.8pt}
  {\vrule height6.3ptdepth-.5ptwidth.8pt\kern-.8pt}
  {\vrule height4.1ptdepth-.35ptwidth.6pt\kern-.6pt}
  {\vrule height3.1ptdepth-.25ptwidth.5pt\kern-.5pt}}
\def\fudge{\mathchoice{}{}{\mkern.5mu}{\mkern.8mu}}
\def\bbc#1#2{{\rm \mkern#2mu\vbar\mkern-#2mu#1}}
\def\bbb#1{{\rm I\mkern-3.5mu #1}}
\def\bba#1#2{{\rm #1\mkern-#2mu\fudge #1}}
\def\bb#1{{\count4=`#1 \advance\count4by-64 \ifcase\count4\or\bba A{11.5}\or
  \bbb B\or\bbc C{5}\or\bbb D\or\bbb E\or\bbb F \or\bbc G{5}\or\bbb H\or
  \bbb I\or\bbc J{3}\or\bbb K\or\bbb L \or\bbb M\or\bbb N\or\bbc O{5} \or
  \bbb P\or\bbc Q{5}\or\bbb R\or\bbc S{4.2}\or\bba T{10.5}\or\bbc U{5}\or
  \bba V{12}\or\bba W{16.5}\or\bba X{11}\or\bba Y{11.7}\or\bba Z{7.5}\fi}}

\newcommand{\tsx}{\tilde \sigma_x}
\newcommand{\tsy}{\tilde \sigma_y}
\newcommand{\tsz}{\tilde \sigma_z}

\newcommand{\RR}{\mbox{${\rm \:  R\!\!\!\! I
\;\;}$}}

\newcommand{\vs}{\vspace{0.25cm}}

\newtheorem{theorem}{Theorem}
\newtheorem{itlemma}{Lemma}[section]
\newtheorem{itproposition}[itlemma]{Proposition}
\newtheorem{itcorollary}[itlemma]{Corollary}
\newtheorem{itremark}[itlemma]{Remark}
\newtheorem{itremarks}[itlemma]{Remarks}
\newtheorem{itdefinition}[itlemma]{Definition}
\newtheorem{itexample}[itlemma]{Example}

\newenvironment{lemma}{\begin{itlemma}\rm}{\end{itlemma}} 
\newenvironment{remark}{\begin{itremark}\rm}{\end{itremark}} 
\newenvironment{remarks}{\begin{itremarks} \rm}{\end{itremarks}}
\newenvironment{corollary}{\begin{itcorollary}\rm}{\end{itcorollary}}
\newenvironment{proposition}{\begin{itproposition}\rm}{\end{itproposition}}
\newenvironment{definition}{\begin{itdefinition}\rm}{\end{itdefinition}}
\newenvironment{example}{\begin{itexample}\rm}{\end{itexample}}
\newenvironment{fact}{\noindent {\em Fact}. \ \ }{\hfill \medskip}
\newenvironment{claim}{\noindent {\em Claim}. \ \ }{\hfill \medskip}
\newcommand{\be}[1]{\begin{equation}\label{#1}}
\newcommand{\ee}{\end{equation}}
\newcommand{\bl}[1]{\begin{lemma}\label{#1}}
\newcommand{\br}[1]{\begin{remark}\label{#1}}
\newcommand{\brs}[1]{\begin{remarks}\label{#1}}
\newcommand{\bt}[1]{\begin{theorem}\label{#1}}
\newcommand{\bd}[1]{\begin{definition}\label{#1}}
\newcommand{\bp}[1]{\begin{proposition}\label{#1}}
\newcommand{\bc}[1]{\begin{corollary}\label{#1}}
\newcommand{\bfact}[1]{\begin{fact}\label{#1}}
\newcommand{\bex}[1]{\begin{example}\label{#1}}
\newcommand{\ec}{\end{corollary}}
\newcommand{\efact}{\end{fact}}
\newcommand{\eex}{\end{example}}
\newcommand{\el}{\end{lemma}}
\newcommand{\er}{\end{remark}}
\newcommand{\ers}{\end{remarks}}
\newcommand{\et}{\end{theorem}}
\newcommand{\ed}{\end{definition}}
\newcommand{\ep}{\end{proposition}}
\newcommand{\epr}{\end{proof}}
\newcommand{\bpr}{\begin{proof}}
\newcommand{\bcl}{\begin{claim}}
\newcommand{\ecl}{\end{claim}}
\newcommand{\bi}{\begin{itemize}}
\newcommand{\ei}{\end{itemize}}
\newcommand{\ben}{\begin{enumerate}}
\newcommand{\een}{\end{enumerate}}

\title{\LARGE \bf{Exact Algebraic Conditions for Indirect Controllability in Quantum
Coherent Feedback Schemes}}

\author{Domenico D'Alessandro\thanks{ Department
of Mathematics, Iowa State University, Ames, Iowa, U.S.A.\ \
Electronic address: daless@iastate.edu}\, Francesca Albertini \thanks{ Dipartimento di Matematica Pura ed
Applicata, Universita' di Padova, Italy, \ \ Electronic address:
albertin@math.unipd.it}\    and \, Raffaele
Romano\thanks{ Department of Mathematics, Iowa State University,
Ames, Iowa, U.S.A.\ \ Electronic address: rromano@iastate.edu}
   \\}

\begin{document}

\maketitle

\begin{abstract}
In coherent quantum feedback control schemes,  a target quantum
system $S$ is put in contact with an auxiliary system $A$ and the
coherent control can directly affect only $A$. The system $S$ is
controlled {\em indirectly} through the interaction with $A$. The
system $S$ is said to be {\it indirectly controllable} if every unitary
transformation can be performed on the state of $S$ with this
scheme. The indirect controllability of $S$ will depend on the {\em
dynamical Lie algebra}  ${\cal L}$ characterizing the dynamics of
the total system $S+A$ and on the initial state of the auxiliary
system $A$. In this paper we describe  this characterization
exactly.

A natural assumption is that the auxiliary system $A$ is {\it minimal} which means that there is no part of $A$ which is uncoupled to $S$, and we denote by $n_A$ the dimension of such a minimal $A$, which we assume to be fully controllable. We show that, if $n_A$  is greater than or equal to $3$,  indirect controllability
of $S$ is verified if and only if complete controllability of the
total system $S+A$ is verified, i.e., ${\cal L}=su(n_Sn_A)$ or ${\cal
L}=u(n_Sn_A)$, where $n_S$ denotes the dimension of the system $S$.
If $n_A=2$, it is possible to have indirect
controllability without having complete controllability. The exact
condition for that to happen is given in terms of a Lie algebra
${\cal L}_S$ which describes the evolution on the system $S$ only. We
prove that  indirect controllability is verified if and only if
${\cal L}_S=u(n_S)$,  and the initial state of
the auxiliary system $A$ is pure.

\end{abstract}

\vs


\vs


\vs

\section{Introduction}

In the paper \cite{LloydCF}, S. Lloyd proposed a scheme for control
of quantum systems where the controller itself was a quantum system
which was affecting the target system via the interaction. This
scheme, named  {\it coherent feedback control},  was later expanded in
several ways (see \cite{Jamesreview} for a recent review) and it is
currently object of intensive research. The consideration of this
scheme motivates the fundamental question of to what extent one can
control a quantum system $S$ indirectly through the interaction with an
auxiliary system $A$. A further motivation comes from the fact that, in many experimental set-ups, the target system is not directly accessible for control or it is not advisable to control it directly as the influence of the environment might become too strong during the experiment, therefore destroying the peculiar (potentially useful) features of quantum dynamics (see, e.g., \cite{MCD1}, \cite{MCD2}).  Controllability studies for systems where only
one subsystem ($A$) can be directly accessed, but another system ($S$) is
the target of control, have  been carried out in several papers (see,
e.g., \cite{Daniel}, \cite{Xaoting}).  However always conditions have
been given so that complete controllability of the whole system
$S+A$ (see Definition \ref{CC} below) is verified. In a recent paper
\cite{conRaf}, a study was started of indirect
controllability (for a precise definition see Definition \ref{IC}
below)  and the case where both system $S$ and $A$ are two
dimensional was treated in detail. It was shown that it is possible
to have indirect controllability of system $S$ without having
complete controllability on the system $S+A$ (while the converse
implication is obvious). It was however shown  later in
\cite{conYao} that if the system $A$ is assumed to be in a
perfectly mixed state at the beginning of the control experiment,
then complete controllability is necessary to have indirect
controllability. In this paper, we solve the general problem to
give exact conditions for indirect controllability for systems $S$
and $A$ of arbitrary dimensions.

It is well known in quantum control theory that the dynamical Lie
algebra ${\cal L}$ generated by the Hamiltonians available for the
evolution of a finite dimensional system of dimension $n$ describes  the set of available evolutions for that system (see e.g., \cite{HT}, \cite{JS}). In particular, if ${\cal
L}=su(n)$ (resp.  ${\cal L}=u(n)$), every special unitary evolution (resp.
 every unitary evolution) is available for the system. This is called the
{\it Lie algebra rank condition}.  It is a result of practical use
as it reduces a problem on a Lie group to a linear algebraic  test, and it suggests the  use of
the theory of Lie algebras and Lie groups as a comprehensive approach to the analysis and control of quantum systems.  In
this paper, we shall use the dynamical Lie algebra to characterize
the indirect controllability properties in indirect control schemes.

The paper is organized as follows. In the next section,
we give the main definitions and state the main
results, while deferring the proofs to the rest of
the paper. Section \ref{auxi} is devoted to some technical lemmas
concerning the general structure of Lie subalgebras of the Lie
algebra $u(n)$. In this section, we also recall some results proved
in other papers by the authors (in particular \cite{conYao} and
\cite{conRaf}) which are used in the proof of the main results. In
section \ref{ngeq3}, we prove the result for the case where the
(minimal) dimension of the auxiliary system $A$,  is greater than or equal to
three. In this case, indirect controllability and complete
controllability are equivalent properties independently of the
initial state of the system $A$. This equivalence does not hold in
the case where this dimension  is $2$. The proof of the indirect controllability
condition in this case is separated in two parts presented in sections \ref{neq2} and \ref{neq2bis}.
This proof is, in fact, quite long and technical and is made up of the treatment of several special cases. In order to streamline the proof we report some of the special cases in appendix B.
We give some concluding remarks in
section \ref{ConRem}.

\section{Basic Definitions and Main Results}
\label{DefandMR}

Although the general definitions of  {\it controllability} for
quantum mechanical systems can be given for systems of infinite
dimension, this property is much better
understood in the finite dimensional case. We shall restrict ourselves to this case, and
denote by  ${\cal{H}}_S$ the Hilbert space of dimension $n_S$  of
 the target system $S$ and by ${\cal{H}}_A$ the Hilbert space of dimension $n_A$ of
the  auxiliary system $A$. The total system
$S+A$ evolves on the  Hilbert space ${\cal{H}}:= {\cal{H}}_S\otimes {\cal{H}}_A$. The dimension of the total system
$S+A$ is $n_{SA}:= n_S \times n_A$. In assigning a dimension $n_A$ to $A$ we are making a natural minimality assumption, namely we assume that $A$ is fully coupled to $S$, that is, it does
not contain any subsystem which is completely decoupled from $S$. It is clear in fact that the dimension of $A$ could be made arbitrarily large by adding `dummy' subsystems or energy levels which are not coupled to $S$.

 Recall that the state of a
quantum mechanical system is described by a {\it density matrix} $\rho$ (see, e.g., \cite{Sakurai}),
i.e., a Hermitian, trace 1, positive semi-definite operator (matrix)
on the Hilbert space associated with the system. We shall denote by
$\rho_S$, $\rho_A$ and $\rho_{TOT}$, the states of the systems $S$,
$A$ and $S+A$, respectively. We also shall make the
assumption that the system $S+A$ has been prepared at the beginning
of the  control experiment in an uncorrelated state, i.e., at time $0$,
 \be{attime0} \rho_{TOT}=\rho_S \otimes \rho_A.  \ee
In typical experimental set-ups, the dynamics of the total system $S+A$
is determined by a set ${\cal F}$ of Hermitian operators on the Hilbert
space ${\cal{H}}$. These are the {\it Hamiltonians} associated with the
system. In the control theory setting, elements in ${\cal F}$ are parametrized by a {\it control variable} $u$ which is allowed to take values in a set ${\cal U}$, so that ${\cal F}:=
\{ H_u  \, | \, u \in {\cal U} \}. $
Thus,
the dynamics of the model is given by
\be{dinamicanuova}
\rho_{TOT}(t) = U(t) \rho_{TOT}(0)U^{\dagger}(t),
\ee
where the unitary operator $U(t)$ is the solution of the {\em Schr\"odinger Operator Equation}:
\be{uevol}
i\dot{U}(t)= H_{ u}  U(t),  \   \  \  \  \  U(0)={\bf{1}}_{n_{SA}}, \footnote{${\bf 1}_q$ (${\bf 0}_q$)  denotes the identity (zero) operator on a Hilbert space of dimension $q$.  We shall often
omit this  subscript when the dimension is obvious from the context. ${\bf 1}_{r,s}$ denotes the square matrix of dimension $r+s$, $\begin{pmatrix}{\bf 1}_r & 0 \cr 0 & -{\bf 1}_s\end{pmatrix}$.}
\ee
and the control parameter $u$ varies with time in the  set ${\cal U}$.   In the Schr\"odinger equation (\ref{uevol}) we have assumed to use units so that the Planck constant $\hbar$ is equal to $1$.
 A typical situation in
experiments is when the $H_u$'s  are linear in $u$, i.e., they have the
form $H_u:=H_0 + \sum_j H_j u_j$ for some finite number of
Hamiltonians $H_0$, $H_j$'s and control variables $u_j$. We shall
assume in the following that all the Hamiltonians involved have zero
trace. This is done without loss of generality because the
introduction of the trace in the Hamiltonians only has the effect of
introducing a phase factor in the evolution of the state which has
no physical meaning.

The controllability of a
finite dimensional quantum system (see, e.g., \cite{Mikobook}, \cite{HT}, \cite{JS})  can be assessed by analyzing the
{\it (dynamical) Lie algebra}  ${\cal L}$ generated by the
Hamiltonians available for the evolution of the system. This is the smallest subalgebra of
$su(n_S n_A)$ containing $i{\cal F}$.\footnote{For a set or space ${\cal F}$ of matrices
we shall often use the notation $i{\cal F}$ to indicate the set or space consisting of the elements in ${\cal F}$ multiplied by the imaginary unit $i$. This allows to go from Hermitian to skew-Hermitian matrices and viceversa.}
If $e^{\cal L}$ denotes the Lie group associated with
${\cal L}$, then the set of possible evolutions for the quantum
system is dense in $e^{\cal L}$ and it is equal to $e^{\cal L}$ if
$e^{\cal L}$ is compact.

\bd{CC} A  quantum system is said to be {\it completely
controllable}  if, for any  special unitary transformation $U_f$,  there
exists a feasible evolution (i.e., a sequence of exponentials of the
form $e^{-iH_ut}$, with $t \geq 0$ and $H_u$ in ${\cal F}$) realizing
that transformation (i.e., whose product is equal to $U_f$). \ed

\bt{LARC} (\cite{HT}, \cite{JS}) A system $S+A$ is completely
controllable if and only if ${\cal L}=su(n_{SA})$. \et

In the case where ${\cal L}$ is only a {\it proper} Lie subalgebra of
$su(n_{SA})$, the knowledge of ${\cal L}$ still gives information on
the dynamics of the system. In particular decompositions of
${\cal L}$ correspond to decompositions of the dynamics of the
system \cite{Mikodec}, \cite{Tannor}, and there exists a fascinating
interplay between symmetries in quantum dynamics and the structure
of the dynamical Lie algebra ${\cal L}$ \cite{Zeier}.

\vs

In the indirect control setting, the Hilbert space associated to the system $S+A$, is the tensor product of the space associated with $S$ and the space associated with $A$. In all operators expressed as tensor product in the following, the operator on the left acts on the Hilbert space associated with  $S$ while the operator on the right acts on the Hilbert space associated with $A$. In this setting, we make the following two assumptions on the dynamics of our model:

\begin{itemize}
\item[{\bf (A-a)}]
The set ${\cal F}$ contains at least one element with nonzero
component on the space of operators
$$\texttt{span} \{ S
\otimes \sigma \, | \, S \in su(n_S), \, \sigma \in su(n_A)\}.$$
\end{itemize}
 This is a natural assumption because it means that there
exists an available Hamiltonian modeling the {\it interaction}
between $S$ and $A$. If that was not the case, then all the operators in ${\cal F}$ would be of the form $F_S\otimes {\bf 1}_{n_A}$ and of the form ${\bf 1}_{n_S} \otimes F_A$,
 and system $S$ and $A$ would evolve independently (all elements in $e^{\cal L}$ would be of the form $U_S \otimes U_A$ (local transformations), with $U_S$ unitary on the system $S$ and $U_A$ unitary on the system $A$.
\begin{itemize}
\item[{\bf (A-b)}]
The dynamical Lie algebra ${\cal L}$ contains all matrices of the
form ${\bf 1}_{n_S} \otimes \sigma$, with $\sigma \in
su(n_A)$.
\end{itemize}
This fact means that  we
have full unitary control on the auxiliary system $A$. Whether this
control is directly available in the experimental set up or it
results from the back-action of the system $S$ on $A$, it is
irrelevant from a mathematical point of view.

We also recall that we have a standing minimality assumption on $A$ in that $n_A$ denotes the dimension of the part of $A$ which is fully coupled with $S$ and does not take into account possibly decoupled additional subsystems.

With initial condition $ \rho_{TOT}= \rho_S \otimes \rho_A $, the
set of available states for $S+A$ is (dense in \footnote{We shall
neglect in the following this distinction and refer to the set
${\cal O}$ as the set of available states for $S+A$. In fact all the
Lie groups we will encounter will be compact so that equality
holds.}) \be{R} {\cal O}:= \{ U \rho_S \otimes \rho_A U^\dagger | U
\in e^{\cal L}\}.  \ee

The set of possible values for $\rho_S$ is obtained by taking the
partial trace with respect to the system $A$ of the elements in
${\cal O}$, i.e., it is the set of matrices  \be{TrAR} Tr_A({\cal
O}):=\{ Tr_A(U\rho_S \otimes \rho_A U^\dagger) \, | \, U \in e^{\cal
L}\}. \ee

The topic of this paper is indirect controllability as described in
the following definition.

\bd{IC} A quantum system $S$ is said to be {\it indirectly
controllable given $\rho_A$}, initial state of the auxiliary system  $A$, if for
every $X \in SU(n_S)$, there exists a reachable evolution $U \in
e^{\cal L}$ of the whole system $S+A$ such that \be{O1} Tr_A(U
\rho_S \otimes \rho_A U^\dagger)=X \rho_S X^\dagger,  \ee  i.e., $X
\rho_S X^\dagger \in Tr_A({\cal O})$, in (\ref{R}), (\ref{TrAR}), for every $\rho_S$, initial
state of $S$. Equivalently, in terms of maps, the system $S$ is indirectly controllable given $\rho_A$, if, for every unitary $X$, there exists $U \in e^{\cal L}$ such that the map $\rho_S \rightarrow Tr_A(U \rho_S \otimes \rho_A U^\dagger)$ coincides with the map $\rho_S \rightarrow X \rho_S X^\dagger$.   \ed

Our goal is to give necessary and sufficient conditions for indirect
controllability given $\rho_A$, in terms of the dynamical Lie
algebra ${\cal L}$ and $\rho_A$ itself. The situation is different if $n_A \geq 3$ and if $n_A=2$. Theorems \ref{Theo1} and
\ref{Theo2} below are our main results.

\bt{Theo1} Assume  $n_A \geq 3$,   and let   $\rho_A$  be any
initial state of the auxiliary system  $A$.
 $S$ is indirectly controllable
given $\rho_A$ if and only if   $S+A$ is   completely controllable,
i.e., ${\cal L}=su(n_{SA})$. \et

\noindent Notice in particular, as a consequence of this result, that for $n_A
\geq 3$ indirect controllability does not depend on the initial
state $\rho_A$ of $A$.

In the  case $n_A=2$, this equivalence is false as shown
in \cite{conFraeasy}, \cite{conRaf}. In order to state the result in
this case,  we consider two subspaces of $su(n_S)$.
We let:
\be{definizioneKP}
\begin{array}{lcl}
{\cal K} & = & \left\{ K\in su(n_S) \ | \ K \otimes {\bf 1}_{n_A}\in {\cal L} \right\} \\
{\cal P} & = & \left\{ P \in su(n_S) \ | \ \exists\,  \sigma_1\in su(n_A), \sigma_1 \not= 0,  \text{ with  } iP \otimes  \sigma_1\in{ \cal L} \right\}.
\end{array}
\ee
Notice that under assumption {\bf (A-b)} ${\cal P}$ contains at
least $i {\bf 1}_{n_S}$. Moreover, it follows from the {\it Simplicity
Lemma} (Lemma 2.2 in \cite{conYao}), under assumption {\bf (A-b)},
that if $iP \otimes \sigma_1 \in {\cal L}$, then $iP \otimes
\sigma \in {\cal L}$ for {\it every} $\sigma \in su(n_A)$. Thus in the definition of ${\cal P}$ above we may
write $\forall$ instead of $\exists$. It also
follows from the {\it Disintegration Lemma} (Lemma 2.3 in
\cite{conYao}), again under assumption  {\bf (A-b)}, that  if an element in
${\cal L}$ contains $iP \otimes \sigma$ as a summand, then $iP
\otimes \sigma$ also belongs to ${\cal L}$.
Therefore ${\cal L}$ is the (direct) sum of ${\cal K} \otimes {\bf 1}_{n_A}$ and $i{\cal P} \otimes su(n_A)$,
i.e.,
\be{sommaYYY}
{\cal L}=\{ {\cal K} \otimes {\bf 1}_{n_A} \} + \{ i{\cal P} \otimes su(n_A) \}.
\ee
We shall denote by
${\cal L}_S$ the subspace (which is in fact a Lie algebra)  \be{Lsdefi}
{\cal L}_S:={\cal K} +
{\cal P}.  \ee
Notice that this definitions hold for any value of $n_A\geq 2$, and we shall, in fact, use it both for the case $n_A \geq 3$ and the case $n_A=2$. In the case $n_A=2$, the space ${\cal L}_S$ is used in the statement of the next theorem   to give the characterization of indirect controllability.

\bt{Theo2} Assume $n_A=2$. System $S$ is indirectly controllable
given $\rho_A$ if and only if one of the following two situations
occurs:
\begin{enumerate}
\item ${\cal L}=su(n_{SA})$, i.e., the
system $S+A$ is completely controllable.

\item $\rho_A$ is a pure state and ${\cal L}_S=u(n_S)$.
\end{enumerate}
\et
\bex{esempioaggiunt}
Consider an Hamiltonian for two spin $\frac{1}{2}$ particles, $S$ and $A$,  interacting via Ising interaction. We assume a constant electro-magnetic field on the spin $S$ and full (time varying electro-magnetic) control on $A$. Such an Hamiltonian may  be given by
\be{jh}
H_u=J\sigma_x \otimes \sigma_x+ i{\bf 1}_2 \otimes \sigma_x u_x(t)+i{\bf 1}_2 \otimes \sigma_y u_y(t)+
\omega_z i\sigma_z \otimes {\bf 1}_2.
\ee
Here $\sigma_{x,y,z}$ are the Pauli matrices defined in (\ref{Paulimat}) below, $J$ the coupling constant, $u_x$ and $u_y$ are the components of the (control) electro-magnetic field in the $x$ and $y$ direction and $\omega_z$ the Larmor frequency. By setting $(u_x, u_y)=(0,0)$, and then $(u_x, u_y)=(1,0)$,  and then $(u_x, u_y)=(0,1)$, we find that the dynamical Lie dynamical ${\cal L}$ contains the matrices $\{iJ\sigma_x \otimes \sigma_x- \omega_z \sigma_z \otimes {\bf 1}_2, {\bf 1}_2 \otimes \sigma_x, {\bf 1} \otimes \sigma_y \}$. Using the commutation and anti-commutation relations for Pauli matrices (see  (\ref{commurel}), (\ref{anticommurel}) below)\footnote{along with (\ref{tenscom}), (\ref{tensantcom})}  the Lie algebra generated by these matrices is given by
\be{Lialgex}
{\cal L}= \texttt{span}\left\{ i \{ \sigma_x, \sigma_y \} \otimes \{\sigma_x, \sigma_y, \sigma_z \}, \, {\bf 1}_2 \otimes \{ \sigma_x, \sigma_y, \sigma_z \} , \, {\sigma_z} \otimes {\bf 1}_2 \right\}.
\ee
This is in fact the dynamical Lie algebra associated with the system since $iH_u$ in (\ref{jh}) for every $u \in \RR$ belongs to ${\cal L}$. A simple dimensions count shows that ${\cal L} \not= su(4)$\footnote{Since $\texttt{dim}({\cal L})=10$ and $\texttt{dim}(su(4))=15$.} Therefore the system is
{\it not} completely controllable. However the subspaces ${\cal K}$ and ${\cal P}$  of Theorem \ref{Theo2} are
 given by ${\cal K}:=\texttt{span} \{ \sigma_z\}$ and ${\cal P}:=\texttt{span} \{ i{\bf 1}_2,  \sigma_x, \sigma_y\}$. Therefore (see (\ref{Lsdefi})) condition 2. of Theorem \ref{Theo2} is verified if the initial state $\rho_A$ of $A$ is a pure state. In this case, system $S$ is indirectly controllable. A more complete analysis of this example along with a constructive algorithm for indirect control is presented in \cite{conFraeasy}.
\eex

The remainder of the paper is devoted to proving Theorems
\ref{Theo1} and \ref{Theo2}.

\section{Preliminary Results}
\label{auxi}

The following Lemma, which was proved in \cite{conRaf} (cf. Theorem 1 in that paper),
is going to be a basic tool to prove the necessity of the
conditions for indirect controllability. Let $\rho_S \otimes \rho_A$
be the initial condition of the system $S+A$, and, given the
dynamical Lie algebra ${\cal L}$,
consider the space\footnote{Recall that, for a Lie algebra ${\cal L}$,
and a vector space of matrices ${\cal M}$,   the space $ad_{\cal L}^k {\cal M}$
is defined recursively as $ad_{\cal L}^0 {\cal M}={\cal M}$,
$ad_{\cal L}^{k+1}=ad_{\cal L} (ad_{\cal L}^k {\cal M})$,
where $ad_{\cal L}{\cal M}$ is the span of all matrices of
the form $[L,M]$, with $L \in {\cal L}$ and $M \in {\cal M}$.}
\be{calVVV}
{\cal V}:= \bigoplus_{k=0}^\infty ad^k_{\cal L} \, \texttt{span} \{ i \rho_S \otimes \rho_A \}.
\ee
We have the following
\bl{Lemmabasic} Given $\rho_S\not=\frac{1}{n_S}{\bf 1}_{n_S}$ and $\rho_A$, assume that for every $X \in SU(n_S)$ there exists a $U \in e^{\cal L}$ such that
\be{ICspec}
Tr_A\left( U \rho_S \otimes \rho_A U^\dagger\right)=X\rho_S X^\dagger.
\ee
Then,
\be{Tracciacondition}
Tr_A ({\cal V})=u(n_S).
\ee
\el
Notice that this necessary condition is given for `non uniform' indirect controllability, which is a weaker property than the one defined in Definition \ref{IC}. This means that the transformation $U$ in (\ref{ICspec}) could, in principle, depend on $\rho_S$. The property (\ref{Tracciacondition}) is therefore also necessary for indirect controllability as in Definition \ref{IC}.    It is known that this condition is, in general,  not sufficient \cite{conRafCDC}.

\vs

We now  recall the definition of the Pauli matrices $\sigma_{x,y,z}$ in quantum mechanics,
\be{Paulimat}
\sigma_x:=\frac{1}{2} \begin{pmatrix} 0 & i \cr i & 0 \end{pmatrix} \qquad
\sigma_y:=\frac{1}{2} \begin{pmatrix} 0 & -1 \cr 1 & 0 \end{pmatrix} \qquad
\sigma_z:=\frac{1}{2} \begin{pmatrix} i & 0 \cr 0 & -i \end{pmatrix},
\ee
which satisfy the commutation relations,
\be{commurel}
[\sigma_x, \sigma_y]=\sigma_z, \qquad [\sigma_y, \sigma_z]=\sigma_x, \qquad [\sigma_z, \sigma_x]=\sigma_y,
\ee
and anti-commutation relations,
\be{anticommurel}
\{ \sigma_{j}, \sigma_{k}\} =-\frac{1}{2} \delta_{j,k}{\bf 1}_2,
\ee
$j,k=x,y,z$.\footnote{In the following, we shall be interested
in commutators and anti-commutators of matrices that are tensor products of two matrices. The
following relations will be repeatedly used without necessarily being explicitly mentioned:
\be{tenscom}
[A \otimes B, C \otimes D]=\frac{1}{2} \left( [A,C] \otimes \{ B,D \} + \{ A ,C \} \otimes [B,D] \right),
\ee
\be{tensantcom}
\{A \otimes B, C \otimes D\}=\frac{1}{2} \left( \{A,C\} \otimes \{B,D\} + [ A,C ] \otimes [B,D] \right).
\ee
}

\vs

Much of the proof of our theorems will be based on
understanding the nature of the subspaces   ${\cal K}$ and
${\cal P}$ defined in equation (\ref{definizioneKP}).
These spaces satisfy the commutation relations
of a Riemannian symmetric space \cite{Helgason}, i.e.,
 \be{Riem}
[{\cal K}, {\cal K}] \subseteq {\cal K},  \qquad [{\cal K}, {\cal
P}] \subseteq{\cal P}, \qquad [{\cal P}, {\cal P}] \subseteq {\cal
K}. \ee The first two of these relations are obvious from the definition, while the
third one is obtained by calculating, given $P_1$ and $P_2$ in
${\cal P}$,
\be{thirdone} -\frac{1}{2} \sum_{j=1}^{n_A}  [iP_1
\otimes \Sigma_j, iP_2 \otimes \Sigma_j]=[P_1, P_2] \otimes {\bf 1}_{n_A}
\in {\cal L}.   \ee
Here $\Sigma_j$, $j=1,\ldots,n_A$,  denotes the
matrix in $su(n_A)$ with $i$ and $-i$ in position $j$ and $j+1 \,
\mod(n_A)$, on the main diagonal, respectively,  and zeros everywhere
else.

\noindent We also have  the anti-commutation relation
 \be{anticommurel2}
i\{ {\cal P}, {\cal P} \} \subseteq {\cal P}.   \ee
In order to see this consider $\sigma_x$ and
$\sigma_y$ the standard Pauli matrices in
$su(2)$ which satisfy the commutation and
anti-commutation relations (\ref{commurel}), (\ref{anticommurel}).
In $su(n_A)$, with $n_A \geq 3$,  we  denote in the following calculation (\ref{contocommurel}), with
some abuse in notation, by $\sigma_{x,y,z}$,  matrices which have
the corresponding Pauli matrix in the diagonal block corresponding to the
first two rows and columns and zeros everywhere else. By extending
naturally the commutation and anti-commutation relations (\ref{commurel}),
(\ref{anticommurel}),  we have for any $P_1$ and $P_2$ in ${\cal P}$
\be{contocommurel} [iP_1 \otimes \sigma_x, iP_2 \otimes
\sigma_y]=\frac{i}{2} i\{P_1,P_2\} \otimes \sigma_z \in {\cal L}.
\ee
From this, equation (\ref{anticommurel2}) follows by definition.\footnote{See the comment on the Simplicity Lemma and the definition of ${\cal P}$ after formula (\ref{definizioneKP}).}



The following two lemmas are the first step to understand the structure of ${\cal P}$.
\bl{Vandermonde} Let ${\cal A}$ be a maximal Abelian subalgebra of ${\cal L}_S$, with ${\cal A} \subseteq {\cal P}$.\footnote{The word {\it maximal} means that it is not a proper subalgebra of any Abelian subalgebra which is also contained in ${\cal P}$.} After a possible change of coordinates on ${\cal L}_S$,\footnote{By a change of coordinates we mean a transformation ${\cal L}_S \rightarrow T {\cal L}_S T^\dagger$, with $T \in U(n_S)$. Such a transformation does not affect the properties of indirect controllability of system $S$.}
a basis of ${\cal A}$ is given  by \be{favbasis} D_1:={\text{diag}} \{ i{\bf
1}_{n_1},{\bf{0}}_{n_2},\ldots, {\bf{0}}_{n_l} \},..., \ee
\[
D_l:={\text{diag}} \{ {\bf 0}_{n_1},{\bf{0}}_{n_2},\ldots,
i{\bf{1}}_{n_l} \},
\]
for some integers $n_1$,...,$n_l$.
\el

\vs

\noindent The proof is given in Appendix A.

\vs
To further investigate  the subspace ${\cal P} \subseteq {\cal L}_S$,  we introduce
a partition of the row and the column indexes and a block structure in the matrices in ${\cal P}$
according to (\ref{favbasis}). Each index  $j=1,\ldots,l$,  corresponds  to a set of indices
of the rows and the columns of matrices in ${\cal P}$, the set being of cardinality $n_j$. Let us introduce
an auxiliary undirected graph ${G}_{\cal P}$ whose nodes correspond to  the indices $\{ 1,2,\ldots, l\}$.
There is an edge between the node $j$ and the node $k$, ($j \not=k$) if and only if there is a
matrix in ${\cal P}$ such that the $(j,k)$-th block (and therefore the $(k,j)-$th block since the matrix is skew-Hermitian) is different from zero.   We have the following Lemma on the structure of ${\cal P}$.

\bl{StructureP}
Let $G_{\cal P}$ be the indirect graph defined above. Then we have:
\begin{enumerate}

\item If $G_{\cal P}$ is not connected there exists a change of coordinates to put all matrices
of ${\cal P}$ in block diagonal form with the $r$-th block,  corresponding to the indices of the $r$-th  connected component of $G_{\cal P}$, $I_r$, having dimension $\sum_{j \in I_r}n_j$.

\item If ${G_{\cal P}}$ is connected then $n_1=n_2=\cdots=n_l$.

\end{enumerate}
\el

\bpr The first statement of the Lemma is a consequence of the
definition of the graph $G_{\cal P}$. Perform  a change of
coordinates which puts together indexes corresponding to the same
connected component of the graph. If $j$ and $k$ are two block
indices corresponding to different components,  each block at the intersection of the $j$-th and $k$-th row and column block for every matrix in ${\cal P}$ is zero, by definition. So the
matrices in ${\cal P}$ have the corresponding block diagonal
structure.

To show the second point of the Lemma denote by $P_{j,k}$ a matrix different from zero at the intersection of the $j$-th and $k$-th row and column block. Let $R_{j,k}$ be the block
different from zero at the intersection of the $j$-th and $k$-th index in $P_{j,k}$ (with the block at the
intersection of the $k$-th and $j$-th position equal to $-R_{j,k}^\dagger$).  Using the basis matrices $D_1,\ldots,D_l$,
defined in (\ref{favbasis}),  we  calculate  $\hat P_{j,k}:=[D_j,[D_k,P_{j,k}]]$ which is in ${\cal P}$ because
of (\ref{Riem}). The matrix $\hat P_{j,k}$ contains only zeros except in the $(j,k)-$th  and $(k,j)-$th  block which are
occupied by $R_{j,k}$ and $-R_{j,k}^\dagger$ respectively. Consider the matrix $\hat P_{j,k} \in {\cal P}$, with $j <k$, as
defined above. Calculating $i\{\hat P_{j,k}, \hat P_{j,k} \} \in {\cal P}$, we see that this matrix is
zero except for the $(j,j)$-th and $(k,k)$-th block that are equal to $-2iR_{j,k} R_{j,k}^\dagger$ and
$-2i R_{j,k}^\dagger R_{j,k}$ respectively. Since this new matrix commutes with the maximal Abelian
algebra in ${\cal A}$ defined in Lemma \ref{Vandermonde},  both matrices must be multiples of the identity
in dimensions $n_j$ and $n_k$, respectively, from which we get
\be{multipli}
R_{j,k} R_{j,k}^\dagger= \alpha  {\bf 1}_{n_j}, \qquad R_{j,k}^\dagger R_{j,k}=\beta  {\bf 1}_{n_k}.
\ee
Since $R_{j,k} \not=0$ both $\alpha$ and $\beta$ must be different from zero, and we have
\be{usoilrango}
n_j=\texttt{rank}(R_{j,k} R_{j,k}^\dagger)=\texttt{rank}(R_{j,k}^\dagger  R_{j,k})=n_k.
\ee
Since the graph $G_{\cal P}$ is connected, taking a path between any two
nodes and repeating this argument between neighboring nodes, it follows that $n_1=n_2=\cdots=n_l$.
\epr

\subsection{Some results on Lie subalgebras of $u(n)$ and symmetric spaces, with application to ${\cal L}_S$}
\label{specres}
In the attempt to understand the nature of ${\cal L}_S$ we shall use some results about general Lie
algebras and, in particular, Lie  subalgebras of $u(n)$. We recall here
these results, refer to standard texts on Lie algebras, Lie
groups and symmetric spaces such as \cite{Helgason} for further
details, and report (in Appendix A) some proofs we were not able to find in the literature.

Given a Lie algebra ${\cal L}$,  a {\it representation} of ${\cal L}$
is a homomorphism $\Phi \, : \, {\cal L} \rightarrow End\left({\cal V} \right)$, i.e.,
a linear map from ${\cal L}$ to the Lie algebra of endomorphisms of
a vector space ${\cal V}$, satisfying for any $A,B \in {\cal L}$,
$\Phi([A,B])=[\Phi(A), \Phi(B)]$. In this equality, with some abuse of notation, the commutator $[\cdot, \cdot]$ on the left hand side is the commutator
in the Lie algebra ${\cal L}$ while the commutator on the right hand side is the standard matrix commutator, i.e., $[A,B]:=AB-BA$. A particular representation is the
{\it adjoint representation}, $A \rightarrow ad_A$, where the space
${\cal V}$ is ${\cal L}$ itself and $ad_A B:=[A,B]$. The {\it Killing form}
$\langle \cdot, \cdot \rangle_K$ on ${\cal L}$ is defined as
\be{Killing} \langle A, B \rangle_K =Tr(ad_A ad_B).  \ee This form is
{\it bilinear} and {\it symmetric} (i.e., $\langle A,B
\rangle_K=\langle B,A \rangle_K$) as well as {\it invariant}, i.e.,
\be{Jacobiplus} \langle [A,B],C \rangle_K= \langle [B,C],A
\rangle_K. \ee Moreover it is invariant under {\it automorphisms}
$\theta$ of the Lie algebra, i.e., one to one and onto homomorphism
of the Lie algebra to itself. This invariance property means that $\langle \theta(A),
\theta(B)\rangle_K =\langle A, B\rangle_K $.

A Lie
algebra ${\cal L}$ is called {\it simple} if it has no ideals except the trivial ones, i.e., ${\cal L}$, and zero, and its dimension is at least two. It is
called {\it semisimple}  if it is the direct sum of simple
ideals.\footnote{That is ${\cal L}={\cal S}_1 +  {\cal S}_2
+ \cdots + {\cal S}_m$, with ideals ${\cal S}_j$'s, with
$[{\cal S}_j, {\cal S}_k]=0$ and ${\cal S}_j \bigcap {\cal S}_k=\{0\}$
if $j\not=k$.} A Lie algebra is called {\it reductive} if it is the direct sum
of a semisimple Lie algebra and an Abelian Lie algebra. Subalgebras of $u(n)$
are always  reductive. The Killing form is a very important tool in the
analysis of Lie algebras. {\it Cartan's criterion}  states that a
Lie algebra $\cal L$ is semisimple if and only if the corresponding
Killing form is nondegenerate.\footnote{This means that the only $X
\in {\cal L}$ such that $\langle X, Y \rangle_K=0,$ for every $Y$ is
 $X=0$.} Another equivalent condition of semi-simplicity is that $[{\cal L}, {\cal L}]={\cal L}$.   For a semisimple Lie subalgebra of $u(n)$, ${\cal L}$, the Killing form is {\it negative definite}  and the corresponding Lie group, $e^{\cal L}$,  is compact. Moreover, see, e.g., \cite{Humphreys} (II 5.1),  if ${\cal I}$ is an ideal of ${\cal L}$, the Killing form of ${\cal I}$ is equal to the restriction to ${\cal I} \times {\cal I}$ of the Killing form on ${\cal L}$.

Let us now consider again the subspace ${\cal L}_S$ of $u(n_S)$ defined
in (\ref{Lsdefi}). From the commutation relations (\ref{Riem}) it follows
that ${\cal L}_S$ is a Lie algebra and, in fact, a Lie subalgebra of $u(n_S)$ and therefore it is reductive.  The following fact will be useful (see Appendix A for the proof).
\bl{decoK}  Assume ${\cal K} \bigcap {\cal P}=\{ 0 \}$. The subalgebra  ${\cal K}$ of ${\cal L}_S$ can be written as
 \be{poi}{\cal K}=[{\cal P}, {\cal P}] + {\cal
R}
\ee where ${\cal R}$ commutes with ${\cal P}$ and it is an ideal in ${\cal K}$ (and therefore in ${\cal L}_S$).
\el
\noindent The next corollary is a  consequence of
Lemma \ref{decoK}.

\bc{corollario1} Assume that  $su(n)$
has a decomposition $su(n)={\cal K} + {\cal P}$, with ${\cal K} \bigcap {\cal P}=\{0\}$ and satisfying conditions (\ref{Riem}). Then
$[{\cal P}, {\cal P}]={\cal K}$. Furthermore $[{\cal K}, {\cal P}]={\cal P}$. \ec
\bpr
Since $su(n)$ is a simple Lie algebra,
it does not have any nontrivial ideal so, in equation (\ref{poi}), the space ${\cal R}$ must be zero, which implies
$[{\cal P}, {\cal P}]={\cal K}$.  The fact  that $[{\cal K}, {\cal P}]={\cal P}$ can be seen as follows. Assume that
$\hat {\cal P}:=[{\cal K}, {\cal P}] \varsubsetneq {\cal P}$ and
define $\hat {\cal L}:= \hat {\cal P} + {\cal K}$, which is a
proper subspace of $su(n)$. We have \be{opla} [su(n), \hat {\cal
L}]=
[{\cal K} + {\cal P}, {\cal K} +
\hat {\cal P}] \subseteq [{\cal K}, {\cal K}] +
[{\cal K}, \hat {\cal P}] +  [{\cal P}, {\cal K}] +[{\cal
P}, \hat {\cal P}]
\subseteq \hat {\cal L},
\ee
it follows that $\hat {\cal L}$ is an ideal in $su(n)$ which
contradicts the fact that $su(n)$ is a simple Lie algebra. \epr

Decompositions $su(n)={\cal K} + {\cal P}$, with ${\cal K} \bigcap {\cal P} =\{ {\bf 0} \}$ with (\ref{Riem}),
are also called {\it Cartan decompositions} of $su(n)$ and correspond to
symmetric spaces of the corresponding Lie group $SU(n)$ \cite{Helgason}. According
to Cartan classification, modulo a change of coordinates,
there are only three types of such decompositions, which
are denoted by ${\bf AI}$, ${\bf AII}$ and ${\bf AIII}$. We shall
use in the following the decompositions ${\bf AI}$ and a special
case of decomposition ${\bf AIII}$. In particular for the
decomposition ${\bf AI}$, ${\cal K}$ is the space of
(skew-Hermitian, zero trace) real matrices, ${\cal R}{e}$, and
${\cal P}$ is the space of (skew-Hermitian, zero trace) purely
imaginary matrices, ${\cal I}{m}$. Therefore  we write \be{AIdeco}
su(n)={\cal R}{e} + {\cal I}{m}. \ee
If we take ${\cal R}{e}$ as ${\cal K}$ and ${\cal I}{m}$ as ${\cal P}$, conditions (\ref{Riem}) are verified. In the ${\bf AIII}$ Cartan
decomposition, we collect two groups of row and column indices and
decompose the matrices in $su(n)$ in terms of matrices that are
block diagonal with respect to this decomposition, ${\cal D}{i}$,
and anti-diagonal with respect with respect to this decomposition,
${\cal A}{n}$. So that we have \be{AIIIdeco} su(n)= {\cal D}{i}
+ {\cal A}{n}.  \ee If we take ${\cal D}{i}$ as ${\cal K}$ and
${\cal A}{n}$ as ${\cal P}$, conditions (\ref{Riem}) are verified.
In this paper, we shall use the special case where the first group
of row and column indexes contains only the first row and column and
the other group contains the remaining indexes. By combining  the
two above Cartan decompositions, we can construct another one, by
defining \be{L1} {\cal L}_1:= \left( {\cal R}{e} \bigcap {\cal A}{n}
\right) + \left( {\cal I}{m} \bigcap {\cal D}{i} \right),\ee
and \be{L2} {\cal L}_2:= \left( {\cal R}{e} \bigcap {\cal D}{i}
\right) + \left( {\cal I}{m} \bigcap {\cal A}{n} \right).\ee It
is easily verified that $su(n)={\cal L}_2 + {\cal L}_1,$ with ${\cal
L}_1$ and ${\cal L}_2$ satisfying the relations (\ref{Riem}) with
${\cal K}={\cal L}_2$ and ${\cal P}={\cal L}_1$, i.e., \be{newRiem}
[{\cal L}_2, {\cal L}_2] \subseteq {\cal L}_2, \qquad [{\cal L}_2,
{\cal L}_1] \subseteq {\cal L}_1, \qquad [{\cal L}_1, {\cal L}_1]
\subseteq {\cal L}_2. \ee One can also readily verify the following
 anti-commutation relations: \be{antiRiem} i\{ {\cal L}_1,
{\cal L}_1 \} \subseteq {\cal L}_1 + \texttt{span} \{ i {\bf 1}_n\} ,
\qquad  i\{ {\cal L}_2, {\cal L}_2 \} \subseteq {\cal L}_1 +
\texttt{span} \{i {\bf 1}_n \}, \qquad i\{ {\cal L}_1, {\cal L}_2 \}
\subseteq {\cal L}_2. \ee

\vs

If ${\cal L}$ is a semisimple Lie subalgebra of $u(n)$, with a Cartan decomposition ${\cal L}={\cal K} + {\cal P}$, ${\cal K}\bigcap{\cal P}= \{0\}$ with ${\cal K}$ and ${\cal P}$ satisfying (\ref{Riem}), Cartan's theorem (cf. \cite{Helgason}) provides a way to parametrize the corresponding Lie group $e^{\cal L}$. In particular, is ${\cal A} \subseteq {\cal P}$ is a maximal Abelian subalgebra of ${\cal P}$, every element $Y \in e^{\cal L}$, can be written as $Y=K_1 e^{\tilde A} K_2$, with $K_1, K_2 \in e^{\cal K}$ and $\tilde A \in {\cal A}$. We shall use this representation of elements in $e^{\cal L}$ several times in the following.

\subsection{Normal vector spaces}
\label{NormalVS}
When analyzing the structure of the Lie algebra ${\cal L}_S$ and in particular using the property (\ref{anticommurel2}), we will have to consider subspaces of the algebra of $n \times n$ complex matrices which satisfy a `normalization' condition.
We will say that  a  vector space ${\cal N}_n$ of $n \times n$ complex matrices over the field of real numbers is {\it normal} if for every pair of matrices $A$ and $B$ in ${\cal N}_n$, it holds:
  \be{relat}
A^\dagger B+ B^\dagger A= B A^\dagger+A B^\dagger = \alpha {\bf
1}_n, \ee for some real number $\alpha$. In particular any matrix $A
\in {\cal N}_n$ is normal since \be{almostunitary} A
A^\dagger=A^\dagger A=\alpha {\bf 1}_n, \ee for some real $\alpha$.

Normal vector spaces can be mapped isomorphically one to the other
by {\it Doubly  Unitary Conjugacy  Transformations} (DUCT)
determined by a pair of unitary matrices ${\bf U}$ and ${\bf V}$ and defined as
$A \in {\cal N}_n \rightarrow {\bf U} A {\bf V}$. Notice in particular that a DUCT
transformation does not modify the defining relation (\ref{relat}).
A normal vector space can be  defined recursively up to a DUCT tranformation, as described in the following Proposition.

\newpage

\bp{classicnorm} Modulo a DUCT transformation, a normal vector space
${\cal N}_n$ of $ n \times n$ matrices is spanned by the following
matrices
\begin{enumerate}
\item ${\bf 0}_n$, or

\item ${\bf 1}_n$ or

\item ${\bf 1}_n$, $i{\bf 1}_{r,s}$ with $r, s \geq 0$
and $r+s=n$, or

\item ${\bf 1}_n$, $i{\bf 1}_{r,s}$, with
$r=s=\frac{n}{2}$ and matrices of the form $C:= \begin{pmatrix} 0 &
C_{1,2} \cr - C_{1,2}^\dagger & 0 \end{pmatrix},$  where the matrices
$C_{1,2}$ span a normal vector space of $\frac{n}{2} \times
\frac{n}{2}$ matrices, ${\cal N}_\frac{n}{2}$.
\end{enumerate}

\bpr
If ${\cal N}_n$ is not zero, consider a matrix
$A\not= 0$ in ${\cal N}_n$. Because of (\ref{almostunitary}) we can replace
$A$ with $\frac{1}{\sqrt{\alpha}} A$ and assume that $A$ is unitary.
Moreover, by applying a DUCT transformation on ${\cal N}_n$, $A \in
{\cal N}_n \rightarrow {\bf U} A {\bf V}$, with ${\bf U}$ equal to the identity and ${\bf V}
:=A^\dagger$, we can assume that $A$ is the identity matrix ${\bf
1}_n$. Using $A={\bf 1}_n$ in (\ref{relat}) we find that the
Hermitian part of every matrix $B \in {\cal N}_n$ is a multiple of the
identity, which means that (modulo a DUCT transformation) ${\cal N}_n$
is spanned by the identity and skew-Hermitian matrices (if any). If
${\cal N}_n$ has dimension $\geq 2$, let us consider a nonzero
skew-Hermitian matrix $B$. We apply a DUCT transformation of a special form with ${\bf V} = {\bf U}^\dagger$ above (that is a Single Unitary Conjugacy Transformation)  $B
\rightarrow {\bf U} B {\bf U}^\dagger$ which does not modify the identity matrix
and diagonalizes $B$. In this new coordinates,
$B={\texttt{diag}}(ia_1,ia_2,\ldots, ia_n)$ and from the fact that
$BB^\dagger$ is a multiple of the identity, it follows that
$a_1^2=a_2^2=\cdots=a_n^2$. By scaling $B$, we can assume that all
of the $a_j$'s are either $1$ or $-1$, so that, modulo a re-ordering
of row and column indexes, $B=i{\bf 1}_{r,s}$. In the special case where $s=0$ and $r=n$, applying relation (\ref{relat})
with a skew-Hermitian $A$ and $B=i{\bf 1}_n$, we see that $A$ must
necessarily be a multiple of $i{\bf 1}_n$. So there is no other skew-Hermitian
matrix in ${\cal N}_n$ except for multiples of $i{\bf 1}_n$, in this case.
If ${\cal N}_n$ has dimension $\geq 3$, we must have that $1\leq r \leq n-1$. We decompose one extra (not a multiple of $i{\bf 1}_{r,s}$) skew-Hermitian matrix $C$ in a basis of ${\cal N}_n$ as \be{decon45}
C:=\begin{pmatrix}C_{1,1} & C_{1,2} \cr -C_{1,2}^\dagger & C_{2,2} \end{pmatrix},  \ee
where $C_{1,1}$ and $C_{2,2}$ are skew-Hermitian and of dimension $r
\times r$ and $s \times s$ respectively. By using (\ref{relat}) with
$A=C$ and $B=i {\bf 1}_{r,s}$ we discover that $-2i C_{1,1} = \alpha
{\bf 1}_r$ and $2iC_{2,2}=\alpha {\bf 1}_s$, so that the block
diagonal part of $C$ is a multiple of $i {\bf 1}_{r,s}$. Therefore,
in the basis of ${\cal N}_n$,  we can take $C$ of the form
(\ref{decon45}) with $C_{1,1}$ and $C_{2,2}$ equal to the $r \times r$ and $s \times s$ zero
matrix, respectively. The possible matrices $C_{12}$ form themselves a vector
space. Moreover take any possible matrix $C$. By applying
(\ref{relat}) with both $A$ and $B$ qual to $C$, we find that
$C_{1,2}C_{1,2}^\dagger=\alpha {\bf 1}_r$ and $C_{1,2}^\dagger
C_{1,2}= \alpha {\bf 1}_s$ for some real number $\alpha$. This shows
that $C_{1,2}$ is either zero or it has full rank and in that case
$r=s$. Therefore the only case where we can have normal vector space
of dimensions $\geq 3$ is when $r=s=\frac{n}{2}$. In particular $n$
must be even. Moreover the space of all matrices $C_{1,2}$ is such
that if we apply (\ref{relat}) with the corresponding matrices $C$
we obtain relation (\ref{relat}) again for matrices of dimension
$\frac{n}{2}$. Therefore the matrices $C_{1,2}$ span  a normal space,
${\cal N}_{\frac{n}{2}}$ ,  of $\frac{n}{2} \times \frac{n}{2}$
matrices. Moreover notice that a DUCT transformation on this space
$A_{\frac{n}{2}} \rightarrow {\bf U} A{\frac{n}{2}} {\bf V}$ can be obtained by
a single  unitary conjugacy  transformation on ${\cal
N}_{n}$ of the form \be{formatrasf} A_n \rightarrow
\begin{pmatrix}{\bf U} & 0 \cr 0 & {\bf V}^\dagger\end{pmatrix}
\begin{pmatrix} 0 & A_{\frac{n}{2}} \cr - A_{\frac{n}{2}}^\dagger &
0 \end{pmatrix} \begin{pmatrix}{\bf U}^\dagger  & 0 \cr 0 &
{\bf V}\end{pmatrix},  \ee which does not affect the first two matrices we
have found in the basis of ${\cal N}_n$. This gives the recursive  construction described in the statement of the theorem.
\epr

Using DUCT transformations, it is always possible to put
the matrices of a basis of ${\cal N}_n$ in a {\it canonical form} in
which all matrices in ${\cal N}_n$ and the following vector spaces
${\cal N}_j$, $j=\frac{n}{2}, \frac{n}{4},\ldots $ obtained with the
above procedure are the identity ${\bf 1}_j$
$j=n,\frac{n}{2},\ldots$  or the matrix $i{\bf 1}_{r,s}$, with
$r_j +s_j=j$, according to the above described algorithm. \ep

\vs

In the following, we shall also be interested in  cases where the normal
vector space of matrices ${\cal N}_n$ is not only a vector space but
also a Lie algebra when equipped with the standard matrix Lie bracket
($[A,B]:=AB-BA$). We first notice that this property
is {\it not}  invariant anymore under DUCT transformation. However it will
be enough for us to consider the case where the basis of ${\cal N}_n$
is in the canonical form described in Proposition \ref{classicnorm}.\footnote{The
normal spaces ${\cal N}_n$ we will  consider are a factor in a tensor product space
and are obtained after a change of coordinates on this space. We shall be able to
assume that this change of coordinates puts ${\cal N}_n$ in canonical
form.} In this case, there is only a finite number of possible
 cases as we shall see  in the following Lemma.

\bl{fewLiealgebras}
Consider a normal vector space ${\cal N}_n$ with a basis in canonical form. If
${\cal N}_n$ is a Lie algebra,  there are only the following possibilities.
\begin{enumerate}
\item ${\cal N}_n= \{ {\bf{0}}\}$.
\item ${\cal N}_n= \texttt{span} \{{\bf 1}_n\}$,
\item ${\cal N}_n= \texttt{span} \{{\bf 1}_n, \, i {{\bf 1}_{r,s}}\}$
\item
 \be{basiscan2} {\cal N}_n=\texttt{span} \left\{ {\bf 1}_n, \, i {{\bf 1}_{\frac{n}{2},\frac{n}{2}}}, \,
 \begin{pmatrix}0 & {\bf 1}_{\frac{n}{2}} \cr - {\bf 1}_{\frac{n}{2}}  & 0 \end{pmatrix} \,
 \begin{pmatrix}0 & i{\bf 1}_{\frac{n}{2}}  \cr i{\bf 1}_{\frac{n}{2}}  & 0 \end{pmatrix}  \right\}
 \ee
\end{enumerate}
\el
\bpr
Cases 1-3 correspond to the first three cases in the construction of Proposition \ref{classicnorm}. The intermediate case between case 3
and 4 is not possible because the Lie bracket between the second and
third term of the right hand side of (\ref{basiscan2}) gives the
fourth term which therefore has to belong to ${\cal N}_n$. However,
as noted in the proof of
Proposition \ref{classicnorm}, the presence of this matrix  implies that no other linearly independent matrix  can be found. Therefore  these four cases are the only admissible ones.
\epr
\br{remar}
In the last case of the above list we can write the basis of ${\cal N}_n$ in terms of the Pauli matrices so that
\be{intermsofPauli}
{\cal N}_n=\texttt{span} \{ {\bf 1}, \, \sigma_z \otimes {\bf 1}_{\frac{n}{2}},
\, \sigma_y \otimes {\bf 1}_{\frac{n}{2}} \, \sigma_x \otimes {\bf 1}_{\frac{n}{2}} \}.
\ee
\er

\section{Proof of Theorem \ref{Theo1}}
\label{ngeq3}

In Theorem \ref{Theo1} we consider the case
$n_A \geq 3$. However the first Lemma holds for any value of $n_A$.

\bl{Lemmaadditional}
Assume ${\cal P}$ in (\ref{Lsdefi}) is Abelian. Then system
$S$ is not indirectly controllable (independently of $\rho_A$).
\el
\bpr
If ${\cal P}$ is Abelian we can  assume that {\it all} the matrices
in ${\cal P}$ are linear combinations of the elements in the basis
(\ref{favbasis}) of Lemma \ref{Vandermonde} and we can take the
basis of ${\cal P}$ as in (\ref{favbasis}). Partition any $K \in
{\cal K}$ according to the partition in the basis of ${\cal P}$. Since $[{\cal K}, {{\cal P}}] \subseteq {\cal P}$, every matrix in
$[{\cal K}, { {\cal P}}]$ must be a linear combination of the
elements in (\ref{favbasis}).  {}From this fact, it is easy to see
that the matrices $K\in {\cal K}$, must have the following block
diagonal structure: \be{2due} K=\left(
\begin{array}{cccc}
                  K_{1,1} &  0 & 0 & 0 \\
                   0  & K_{2,2} & 0 & 0 \\
                   \vdots&   \vdots& \vdots& \vdots\\
                   0& 0& 0& K_{l,l}\end{array} \right),
\ee with $K_{j,j}\in su(n_j)$. Matrices in the  Lie Algebra ${\cal L}_S$ also have this
block diagonal structure\footnote{Notice that this structure assumes  a particular system
of coordinates but the transformation to get
in these coordinates is a local transformation acting on $S$ only. So, it does
not affect the indirect controllability properties of system $S$.}  and matrices in  ${\cal L}$ also
have  a block diagonal structure induced by this structure. Thus a
matrix $U\in e^{\cal L}$ is  of the
form:\be{2tre} U=\left( \begin{array}{cccc}
                 U_1 &  0 & 0 & 0 \\
                   0  & U_2 & 0 & 0 \\
                   \vdots&   \vdots& \vdots& \vdots\\
                   0& 0& 0& U_l\end{array} \right),
\ee where the blocks $U_j$, $j=1,\ldots,l$  have  dimension
$n_jn_A$.\footnote{Recall that $l\geq 2$ since ${\cal P}$ is not the
span of multiples of the identity because of assumption ({\bf
A-a}).It has dimension at least $2$ and it contains multiples of the identity.}
Choose any initial state $\rho_S \otimes \rho_A$ where
$\rho_S$ has the same block structure as in equation (\ref{2due}).
Then for any $U\in e^{\cal L}$, the matrix  $\text{Tr}_A \left(
U\rho_S\otimes \rho_A U^{\dag} \right)$ will have  the same  block
diagonal structure as in equation (\ref{2due}). Since not all the
matrices unitarily equivalent to $\rho_S$ have this diagonal
structure, the model is not indirectly controllable.
\epr

\bl{primo-lemma} Assume that $n_A\geq 3$ and let  $\rho_A$  be
any initial state of the auxiliary system $A$. If $S$ is indirectly
controllable given $\rho_A$ then ${\cal L}_S=u(n_S)$ \el

\bpr  Assume that $S$ is not indirectly controllable and assume by contradiction, that ${\cal L}_S \neq u(n_S)$. If ${\cal P}$ is Abelian,  from the previous Lemma we already know that indirect controllability is not verified. Therefore we can assume that ${\cal P}$ is not Abelian. We prove  the Lemma in two steps.
\begin{itemize}
\item[(a)] If ${\cal P}$ is not Abelian then ${\cal K} \cap {\cal P} \not=0$.
\item[(b)] If  ${\cal K} \cap {\cal P} \not= \{0\}$ and  ${\cal L}_S \not= u(n_S)$ then indirect controllability is not verified.
\end{itemize}
 For the   step (a)  assume ${\cal P}$ is not Abelian and
let $P_1, \, P_2 \in {\cal P}$ such that $[P_1, P_2]=K \neq 0$, with
$K\in {\cal K}$. Since $n_A \geq 3$ there
exist $\sigma_1, \, \sigma_2\in su(n_A)$ such that
 \be{3quattro}
 \ [\sigma_1,\sigma_2]=0, \ \ \ \{\sigma_1,\sigma_2\}= {\bf 1}+i\hat{\sigma},\ee
 with ${\hat{\sigma}}\in su(n_A)$, different from zero.  We have
\be{3unobis}
 \begin{array}{ll}
 [iP_1\otimes\sigma_1, iP_2 \otimes\sigma_2] &= -1/2 [P_1,P_2] \otimes ({\bf 1} +i{\hat{\sigma}}) \\
 & =
 -1/2 K \otimes ({\bf 1} +i{\hat{\sigma}})\in {\cal L}.  \end{array}
 \ee
Since $K\otimes   {\bf 1}\in {\cal L}$, it follows that $iK\otimes
\hat\sigma\in {\cal L}$. Thus $K \in {\cal K} \bigcap {\cal P}$, which shows that ${\cal K} \cap {\cal P} \not= \{ 0 \}$.
Now we show  part (b). It will follow from the proof that part (b)
holds for any value of $n_A$. If ${\cal K}\cap {\cal P} \neq \{0\}$, then given any matrix $B
\not=0$ such that  $B \in {\cal K}\cap {\cal P}$  we choose as
initial state $\rho_S= \frac{1}{n_S}{\bf 1}+\alpha iB$, with $\alpha
\not=0$ and   sufficiently small so that $\rho_S$ is  an admissible
density matrix.\footnote{Note that $B$ cannot be a multiple of the identity because we have assumed at the beginning that all Hamiltonians involved in the dynamics have zero trace, that is, ${\cal L}$ is a subalgebra of $su(n_S n_A)$.} Given any $\rho_A=\frac{1}{n_A}{\bf 1}+i\sigma$, for
$\sigma \in su(n_A)$, we have that $i\rho_S\otimes\rho_A$ belongs to
$\tilde{\cal L}=\text{span}\{i {\bf 1}\otimes{\bf 1}\}
+ {\cal L}$, which is invariant under ${\cal L}$. Therefore
${\cal V}$ defined in (\ref{calVVV}) of
Lemma \ref{Lemmabasic} is such that ${\cal V} \subseteq \tilde {\cal L}$, and we have $Tr_A({\cal V})\subseteq Tr_A(\tilde {\cal L}) ={\cal K}+ \texttt{span} \{ i {\bf 1}\} \subseteq {\cal L}_S \varsubsetneq u(n_S)$, which contradicts Lemma \ref{Lemmabasic}.
\epr

\vs

\noindent The proof of Theorem \ref{Theo1} is now a consequence of
the previous two Lemmas.

\vs

\noindent {\bf Proof of the Theorem}

\vs

We only need to prove that indirect controllability (for a fixed
$\rho_A$) implies ${\cal L}=su(n_S n_A)$. The converse implication
is obvious. Assume indirect controllability. From Lemmas
\ref{Lemmaadditional} and  \ref{primo-lemma}, we know that ${\cal
L}_S=u(n_S)$. Let $\tilde {\cal P}$ the subspace of matrices in
${\cal P}$ with zero trace. We will establish that
 all the matrices of the type $iK  \otimes \sigma$ and $P\otimes {\bf{1}}$,
with $K\in {\cal K}$, $P \in \tilde {\cal P}$ and $\sigma \in
su(n_A)$, are in ${\cal L}$. This implies that ${\cal L} =su(n_S
n_A)$. Notice that since $\tilde {\cal L}_S:= {\cal K} + \tilde
{\cal P}=su(n_S)$, from Corollary  \ref{corollario1} (or
from Lemma \ref{Lemmaadditional}),
it follows that
$\tilde {\cal P}$ cannot be Abelian.

 Since $n_A\geq 3$, we may take    $\sigma_1, \, \sigma_2\in su(n_A)$ such that equation
  (\ref{3quattro}) is satisfied. Then,  as computed in  equation (\ref{3unobis}),
  given $P_1,P_2 \in {\cal P}$, we have:
 \be{3uno}
  [iP_1\otimes\sigma_1, iP_2 \otimes\sigma_2] = -1/2 K \otimes ({\bf 1} +i{\hat{\sigma}}),
 \ee
for $K\in {\cal K}$.
 We can assume $K \not=0$ since $\tilde {\cal
P}$ cannot be Abelian.  In fact, since $[\tilde {\cal P},\tilde
{\cal P}] ={\cal K}$ from Corollary \ref{corollario1}, we have
\[
i K\otimes {\hat{ \sigma}} \in {\cal L}, \ \text {for all } K\in {\cal K}.
\]
Since ${\bf 1}\otimes su(n_A)\in {\cal L}$, from the previous
equation we get that\footnote{From the simplicity Lemma in
\cite{conYao}.} \be{3due}
  i K\otimes   \sigma   \in {\cal L}, \ \text {for all } K\in {\cal K}, \
  \text{ and } \sigma \in su(n_A).
\ee Now calculate \be{thirdonebis} -\frac{1}{2} \sum_{j=1}^{n_A} [iK
\otimes \sigma_j, iP \otimes \sigma_j]=[K, P] \otimes {\bf 1} \in
{\cal L}.   \ee Here $\sigma_j$, $j=1,\ldots,n_A$, denotes the
matrix in $su(n_A)$ with $i$ and $-i$ in position $j$ and $j+1 \,
\mod \,( n_A)$, on the main diagonal, respectively,  and zeros everywhere
else, while $K$ and $P$ are general matrices in ${\cal K}$ and
$\tilde {\cal P}$. Since $[{\cal K}, \tilde {\cal P}]=\tilde {\cal
P}$ from Corollary \ref{corollario1}, we have : \be{3tre} P\otimes
{\bf{1}} \in {\cal L}, \ \text {for all }P\in \tilde {\cal P}. \ee
{}From equations (\ref{3due}) and (\ref{3tre}), the statement
follows. \qed

\section{Proof of Theorem \ref{Theo2}; Part I}
\label{neq2}

From this point on, $n_A=2$.
In the next subsection we prove sufficiency of  conditions $1$ and $2$ of Theorem \ref{Theo2}. In fact,  being condition $1$ obviously sufficient we need to treat only the sufficiency of condition $2$.  Then the proof of necessity is divided in two parts: one in subsection \ref{purenec} and one in section \ref{neq2bis}. Much of the proof of necessity is carried out by looking at the various possibilities for the Lie algebra ${\cal L}_S$. From this analysis there are several special cases to be treated. Some special cases are presented  in Appendix B.

\subsection{${\cal L}_S = u(n_S)$ and $\rho_A$ pure imply  indirect controllability of
$S$}
\bpr
The argument is a generalization to $n_S\geq 2$ of the one
given in \cite{conRaf}.   Assume that
 we want to steer any $\rho_S$ to the unitarily equivalent $X\rho_SX^\dagger$, with $X \in SU(n_S)$, i.e., we need to find a reachable evolution $U\in e^{\cal L}$, such that
 \be{55}
  Tr_A(U \rho_S \otimes \rho_A U^\dagger)=X\rho_SX^\dagger,
  \ee
for every $\rho_S$. Since ${\cal L}_S=u(n_S)$, if we define $\tilde {\cal P}$ the
subspace of ${\cal P}$ of matrices with zero trace, we have from (\ref{Lsdefi})
$su(n_S)={\cal K} + \tilde {\cal P}$, where ${\cal K}$ and
$\tilde {\cal P}$ provide a Cartan decomposition of $su(n_S)$ (see
 (\ref{Riem}) with ${\tilde {\cal P}}$ replacing ${ {\cal P}}$).\footnote{Notice that ${\cal K} \bigcap \tilde {\cal P}=\{0\}$ because if this was not the case (from (\ref{Riem})) ${\cal K} \bigcap \tilde {\cal P}$ would be an ideal of $su(n_S)$ which is excluded since $su(n_S)$ is simple, unless $su(n_S)={\cal K}=\tilde {\cal P}$ which would imply ${\cal L}=su(n_S n_A)$ which gives complete controllability and therefore obviously indirect controllability.} Thus, we can write $X$ as
 \be{Xdef3333}
 X =K_1e^{\tilde A} K_2,
 \ee
where $K_{1,2}\in e^{\cal K}$ and $\tilde A\in {\cal A}$, where ${\cal A}$ is a
maximal Abelian subalgebra (Cartan subalgebra) in $\tilde {\cal P}$
\cite{Helgason} (cf. the discussion at the end of subsection \ref{specres}).

Let
$ \bar{\sigma}:=\left( \begin{array}{cc} -i & 0 \\
                    0& i \end{array} \right).$
The Lie group $e^{\cal L}$ contains all elements of the form: $ K \otimes {\bf 1}, \ {\bf 1}\otimes B, \  e^{it\tilde A\otimes
{\bar{\sigma}}}, $ with $K\in e^{\cal K}$, $B\in SU(2)$ and $\tilde A\in
{\cal A}$. Since  $\rho_A$  is a pure state, there exists a unitary $T$ such that
$T \rho_A T^\dagger= E_1$ where,
$E_1$ is the $2 \times 2$ matrix with $1$ in the $(1,1)$ position and zero
elsewhere. With this $T$, we choose $U \in e^{\cal L}$ given by (cf (\ref{Xdef3333}))
\be{choiceU}
U:=(K_1 \otimes {\bf 1}) (e^{i \tilde A \otimes \bar \sigma})( K_2 \otimes {\bf 1})({\bf 1}\otimes T).
\ee
We verify that
\be{wcU}
U\left(\rho_S \otimes \rho_A\right)
U^{\dag}=
X\rho_S X^\dagger\otimes E_1.
\ee
This follows from the definitions of $U$ and $X$ in (\ref{choiceU}) and (\ref{Xdef3333}) and from the observation that  since $i\bar{\sigma}E_1=iE_1\bar{\sigma}=E_1$,
for a general matrix $\rho$ we have: \be{due}
e^{it \tilde A\otimes \bar{\sigma}}\left(\rho \otimes
E_1\right)e^{-it \tilde A\otimes \bar{\sigma}}= \left(e^{t \tilde A}\rho
e^{-t \tilde A}\right)\otimes E_1. \ee

Taking the partial trace with respect to the system $A$ of (\ref{wcU}) we get equation
(\ref{55}), as desired. \epr

\subsection{If ${\cal L}\not=su(n_S n_A)$, $\rho_A$ pure
is necessary for indirect controllability}
\label{purenec}

We use the main result of \cite{conYao}, i.e., the following
theorem (which we state for $n_A=2$).

\bt{fromSCL} Assume $\rho_A=\frac{1}{2}{\bf 1}$. If for each $X \in
SU(n_S)$ there exists $U \in e^{\cal L}$ which verifies (\ref{55})  for every
density matrix $\rho_S$, then ${\cal L}=su(n_S n_A)$, i.e., complete
controllability is verified.
\et
In other terms, indirect controllability with $\rho_A=\frac{1}{2}{\bf 1}$ implies complete controllability.

Assume that ${\cal L}\not= su(n_Sn_A)$ and $\rho_A$ has the
property that for each $X \in SU(n_S)$ there exists $U \in e^{\cal
L}$ with (\ref{55}) for every density matrix $\rho_S$. From the Theorem \ref{fromSCL}, it follows that $\rho_A$ cannot be the perfectly mixed state, i.e., $\rho_A \not=\frac{1}{2} {\bf 1}$. We want
to prove that $\rho_A$ is necessarily a pure state. Assume this is
not the case. Therefore, $\rho_A = c_1 \rho_{A,1} + c_2 \rho_{A,2}$,
with $c_1 > 0$, $c_2 > 0$, $c_1 + c_2 = 1$ and $\rho_{A,1}$ and
$\rho_{A,2}$ are two projection matrices with $\rho_{A,1} +
\rho_{A,2} = {\bf 1}_2$. From (\ref{55}) we have \be{rafTheo}
c_1 \gamma_1 [\rho_S] + c_2 \gamma_2 [\rho_S] = X \rho_S X^\dagger,
\ee where we have used the definitions of the two trace-preserving completely positive maps (cf., e.g.,
\cite{Petruccione}) $\gamma_1$ and $\gamma_2$,
$\gamma_1 [\rho_S] := Tr_A ( U
\rho_S \otimes \rho_{A,1}  U^\dagger)$ and $\gamma_2 [\rho_S] :=
Tr_A (U \rho_S \otimes \rho_{A,2} U^\dagger)$. Their convex combination can be a unitary map
if and only if both of them realize the same unitary transformation,
that is, for every $\rho_S$, \be{E1} \gamma_1 [\rho_S] = \gamma_2
[\rho_S] = X  \rho_S  X^\dagger. \ee This follows from  the {\it
Choi-Jamiolkowski isomorphism} between trace-preserving completely
positive maps and states~\cite{jamiol}. According to this
isomorphism, given a trace-preserving completely positive map
$\gamma$, acting on the density operators on the Hilbert space
$\mathcal{H}$, the corresponding state is a density operator
$\Gamma$ acting on the space $\mathcal{H} \otimes \mathcal{H}$.  For
our purposes, it is not necessary to describe the exact form of
$\Gamma$.\footnote{$\Gamma$ has the form  $\Gamma := ({\bf 1} \otimes {\gamma}) \, \rho_0,$
where $\rho_0$ is a given  maximally entangled state in $\mathcal{H}
\otimes \mathcal{H}$, and ${\bf 1}$ is the identity operator. In
other words, in the Choi-Jamiolkowski isomorphism, the state
$\Gamma$ associated to the map $\gamma$ is obtained by
acting with
$\gamma$ on a single subsystem of a maximally entangled
pair. More details can be found in \cite{jamiol}.}  This can be
found, along with the proof of the  one-to-one
correspondence between $\gamma$ and $\Gamma$  in~\cite{jamiol}.
In~\cite{jamiol} it is also shown that there is a one-to-one
correspondence between unitary maps acting on the space of
density matrices (a special case of completely positive maps) and pure states in
$\mathcal{H} \otimes \mathcal{H}$. Therefore, the state
corresponding to the unitary transformation in the r.h.s. of
(\ref{rafTheo}), denoted here by $\Gamma_X$, is a pure state. If we
call $\Gamma_1$ and $\Gamma_2$ the states corresponding to
$\gamma_1$ and $\gamma_2$ in (\ref{rafTheo}), we have, because of
the isomorphism, \be{vhi} \Gamma_X=c_1 \Gamma_1 + c_2 \Gamma_2. \ee
Since $\Gamma_X$ is pure this implies  $\Gamma_1 = \Gamma_2 =
\Gamma_X, $ which, from the isomorphism, implies (\ref{E1}). From
(\ref{E1}) we obtain \be{GIH} \frac{1}{2} (\gamma_1[\rho_S] +
\gamma_2[\rho_S]) = \frac{1}{2} Tr_A\left(U \rho_S \otimes
\rho_{A,1} U^\dagger\right)+ \frac{1}{2} Tr_A\left(U \rho_S \otimes
\rho_{A,2} U^\dagger\right)= \ee $$Tr_A\left(U \rho_S \otimes
(\frac{1}{2} {\bf 1})  U^\dagger\right)= X \rho_S X^\dagger,$$ for
every $\rho_S$. Therefore $\frac{1}{2} {\bf 1}$ has the indirect
controllability property. However, this, from Theorem \ref{fromSCL}
implies ${\cal L}=su(n_S n_A)$ which is not verified. Therefore, if
${\cal L}\not=su(n_S n_A)$ the only possibility to have indirect
controllability given $\rho_A$, is when $\rho_A$ is a {\it pure
state}.

\section{Proof of Theorem \ref{Theo2}; Part II: If the system is indirectly controllable given $\rho_A$, then ${\cal L}_S=u(n_S)$}
\label{neq2bis}

This is the longest part of the proof. We have to analyze the Lie
algebra ${\cal L}_S$ in (\ref{Lsdefi}) under the assumption that there is indirect
controllability given $\rho_A$. We know that ${\cal L}_S$ is a
subalgebra of $u(n_S)$ and therefore it is a {\it reductive} Lie
algebra.

We can  assume ${\cal K} \bigcap  {\cal P}= \{
{\bf 0} \}$.
In fact, if ${\cal K} \bigcap  {\cal P} \not= \{
{\bf 0} \}$ then   indirect controllability implies ${\cal L}_S=u(n_S)$ from  statement (b) in the proof of Lemma
\ref{primo-lemma}, which is independent of the assumption $n_A \geq
3$.

We can also assume that the graph $G_{\cal P}$ of
Lemma \ref{StructureP} is connected. If this is not the case, in
appropriate coordinates,  ${\cal P}$ will have a block diagonal form
and ${\cal K}$ will have a corresponding block diagonal form.\footnote{Any
element with nonzero off diagonal in ${\cal K}$ would  give from
$[{\cal K}, {\cal P}] \subseteq {\cal P}$ a corresponding element
with nonzero off-diagonal in ${\cal P}$, using Lie brackets with
elements in the basis (\ref{favbasis}) of Lemma \ref{Vandermonde}.}
This structure is incompatible with the assumption of indirect
controllability because it implies a corresponding block diagonal
structure on the matrices in ${\cal L}$. If $\rho_S$ is chosen
having this block-diagonal structure, this structure will be
preserved after evolution and partial trace. Therefore, $\rho_S$
cannot be transformed in every  matrix which is unitarily equivalent to
itself. A similar argument was used in the proof of Lemma
\ref{Lemmaadditional} to show that ${\cal P}$ cannot be Abelian and
the same argument which was independent of the dimension $n_A$ shows
that ${\cal P}$ cannot be Abelian in this case either.

Given that
the graph $G_{\cal P}$ is connected, we denote by $n_0$ the
dimension $n_1=n_2=\cdots=n_l$ of Lemma \ref{StructureP} and
Lemma \ref{Vandermonde}. Notice that $l$ is always $\geq 2$. $l=1$
would mean that ${\cal P}$ only contains the identity which is incompatible with the Assumption ${\bf (A-a)}$ since in this case there would be no interaction between $S$ and $A$.

Our next
task is to study the possible structure of ${\cal P}$ under the
above conditions, i.e., ${\cal P}$ not Abelian, ${\cal K} \bigcap {\cal P} =\{ {\bf 0} \}$, and $G_{\cal P}$ of Lemma \ref{StructureP} connected.

\subsection{Structure of ${\cal P}$}

We go back to the proof of Lemma \ref{StructureP} and the graph
$G_{\cal P}$, and notice that ${\cal P}$ is spanned by the matrices
$D_1,\ldots,D_l$ (with $n_1=\cdots=n_l=n_0$) in (\ref{favbasis}) of
Lemma \ref{Vandermonde} as well as matrices $P_{j,k}$ ($j<k$, $j,k
\in \{1,\ldots,l\}$) which are zero in every block except for  the
$(j,k)$-th  (and $(k,j)$-th) block. These blocks are occupied by matrices
$R_{j,k}$ (and $-R_{j,k}^\dagger$) which are different from zero and
in fact nonsingular for all pairs $j<k$ for which there is an edge
in the graph $G_{\cal P}$. In fact, there is a nonzero $P_{j,k}$ {\it
for every} pair $j<k \in \{1,\ldots,l\}$. In order to see this, fix
$j$ and $k$ and, since ${G}_{\cal P}$ is connected, fix  a path joining $j$
and $k$. Let $j_0$, $j_1$ and $j_2$ three consecutive nodes on this
path so that there is an edge connecting $j_0$ and $j_1$ and an edge
connecting $j_1$ and $j_2$. Taking the Lie bracket $[D_{j_0},
[P_{j_0,j_1}, P_{j_1,j_2}]]$ which is in ${\cal P}$ we obtain a
matrix which has a nonzero (and nonsingular) block in the position
$(j_0,j_2)$ (and $(j_2,j_0)$). Therefore $j_0$ and $j_2$ also are
connected in $G_{\cal P}$. Repeating this argument and by induction
we see that $j$ and $k$ are also connected.  Therefore, in ${\cal
P}$ there exists an element with all blocks zero except the
$(j,k)$-th and $(k,j)$-th, for every $j<k$. Now denote by ${\cal N}_{j,k}$ ($j<k$,
$j,k \in \{1,2,\ldots,l\}$) the space of $n_0 \times n_0$ matrices
occupying the $(j,k)-$th positions in the matrices in ${\cal P}$.
Because of property (\ref{anticommurel2}), for every pair $(j,k)$,
${\cal N}_{j,k}$ forms a {\it normal vector space} because any two
matrices $A$ and $B$ in ${\cal N}_{j,k}$ satisfy the defining
property (\ref{relat}). The following property considerably
simplifies our analysis.

\bp{Semplificazione} Modulo a change of coordinates on ${\cal L}_S$,
\be{N1etal}  {\cal N}_{1,2}= {\cal N}_{1,3}=\cdots={\cal
N}_{1,l}:={\cal N}_{n_0} \ee and if $j \not=1$, $j < k$, ${\cal N}_{j,k}=i{\cal
N}_{n_0}$. Here ${\cal N}_{n_0}$ (and therefore $i{\cal N}_{n_0}$) is a normal vector space of $n_0 \times n_0$ matrices.
\ep

\bpr Consider given normal matrices $R_{1,k} \in {\cal N}_{1,k}$,
$k=2,\ldots,l$, satisfying (cf. formula (\ref{almostunitary})) $R_{1,k}R_{1,k}^\dagger=R_{1,k}^\dagger R_{1,k}=\alpha_{1,k} {\bf 1}_{n_0}$. After
re-normalization $R_{1,k} \rightarrow \frac{1}{\sqrt{\alpha_1,k}}
R_{1,k}$ (recall that ${\cal N}_{1,k}$ is a vector space), we can
assume that $R_{1,k}$'s are unitary. Perform a change of coordinates
on ${\cal L}_S$ and therefore ${\cal P}$, ${\cal L}_S \rightarrow T
{\cal L}_S T^\dagger$ with $T=\texttt{diag}({\bf 1}_{n_0},
R_{1,2}^\dagger,\ldots,R_{1,l}^\dagger)$, so that the matrix ${\bf 1}_{n_0}$
belongs to each of the ${\cal N}_{1,k}$. This corresponds to DUCT
transformations (cf. subsection \ref{NormalVS}) on the subspaces ${\cal N}_{1,k}$. Every ${\cal
N}_{1,k}$ space is spanned by the identity and possibly (because of
formula (\ref{almostunitary})) by skew-Hermitian matrices. Let us
denote by $E_{1,k}$ the matrix in ${\cal P}$ with the identity ${\bf
1}_{n_0}$ in the $(1,k)-$th block (and $-{\bf 1}_{n_0}$ in the
$(k,1)-$th block) and zeros everywhere else. Let $R_{1,j}$ be a
matrix in ${\cal N}_{1,j}$ which we can assume skew-Hermitian, and
let $\hat R_{1,j}$ the corresponding matrix in ${\cal P}$ which has
zero blocks everywhere except in the blocks $(1,j)$-th and
$(j,1)$-th which are occupied by $R_{1,j}$ and
$-R_{1,j}^\dagger=R_{j,1}$, respectively. By calculating
$[E_{1,j},[E_{1,k}, \hat R_{1,j}]] \in {\cal P}$ we obtain a matrix
which has zeros in every block except for the $(1,k)$-th block which
is (proportional to) $R_{1,j}$ (and accordingly for the $(k,1)-$th
block). This shows ${\cal N}_{1,j} \subseteq {\cal N}_{1,k}$, and
since $j$ and $k$ are arbitrary, equality holds for all $j$ and
$k$'s. Now using the definitions in (\ref{favbasis}), we calculate,
for {\it a given}  $R_{1,j} \in {\cal N}_{1,j}$, and
corresponding $\hat R_{1,j} \in {\cal P}$,
$[D_j, [E_{1,k}, \hat
R_{1,j}]] \in {\cal P}$. This  gives a matrix which has zeros in all blocks
except $iR_{1,j}^\dagger$ in the $(j,k)$-th position (and
accordingly in the $(k,j)$-th position. This shows that $i{\cal
N}_{1,j} \subseteq {\cal N}_{j,k}$. To show the converse inclusion,
calculate $[[\hat R_{j,k}, E_{1,k}],D_1] \in {\cal P}$.
\epr

It follows from the proof of the previous proposition that a change
of coordinates on the Lie algebra ${\cal L}_S$ can be performed in
order to achieve a DUCT transformation on the normal space ${\cal N}_{n_0}$ to put it in the
canonical form described in Proposition \ref{classicnorm}. We shall assume
this to be the case in the following. Let $\tilde {\cal N}_{n_0}$ be the subspace of ${\cal N}_{n_0}$ of skew-Hermitian matrices.
Therefore ${\cal N}_{n_0}=\tilde {\cal N}_{n_0} + \, \texttt{span} \{  {\bf 1}_{n_0}\}$. From
 Proposition \ref{Semplificazione},  we know that a basis for ${\cal P}$ can
 be taken as made up of  the following
 \bi
 \item[1)]
 The matrices $D_1,\ldots,D_l$ in (\ref{favbasis}) and the matrices that  have
 $i{\bf 1}_{n_0}$ in blocks $(j,k)$ (and $(k,j)$) with $j,k=2,\ldots l$ $j<k$;
 \item[2)]
 The matrices
 which have the identity ${\bf 1}_{n_0}$ in the blocks corresponding to the first row (and $-{\bf 1}_{n_0}$ in the blocks corresponding to the  first
 column) (except the diagonal block);
 \item[3)]
 The matrices which have elements
 in a basis of $\tilde {\cal N}_{n_0}$ in the blocks corresponding to the first row (and first column) (except
 the diagonal block);
  \item[4)]
   The matrices which have elements in a basis of $i\tilde {\cal N}_{n_0}$ in
 blocks $(j,k)$ (and accordingly in $(k,j)$),  with $j,k=2,\ldots l$, $j<k$.
 \ei

 This basis can be conveniently expressed using the subspaces defined in (\ref{AIdeco})-(\ref{L2}),
 considering Cartan decompositions of $su(l)$.\footnote{Recall that, in the Cartan decomposition {\bf AIII}, we have chosen to partition the matrices of $su(l)$ in block diagonal and anti-diagonal parts so that the diagonal blocks have dimensions $1 \times 1$ and $(l-1) \times (l-1)$.} In particular:
 The matrices of point 1) above are the
 matrices of $({\cal I}{m} \bigcap {\cal D}{i}) \otimes {\bf 1}_{n_0}
 +\texttt{span}\{ i {\bf 1}_{l} \otimes {\bf 1}_{n_0} \}$; The matrices
 of point 2) are the ones in $({\cal R}{e} \bigcap {\cal A}{n}) \otimes {\bf 1}_{n_0}$; The matrices
 of point 3) are the ones in $i({\cal I}{m} \bigcap {\cal A}{n}) \otimes \tilde {\cal N}_{n_0}$; The matrices
 of the point 4) are the ones in $i( {\cal R}{e}\bigcap {\cal D}{i}) \otimes \tilde {\cal N}_{n_0}$. Therefore, we have
 \bl{espressioP}  With
 the definitions (\ref{L1}), (\ref{L2}),
 \be{s}
 {\cal P}=\left( {\cal L}_1 \otimes {\bf 1}_{n_0} \right)
 + \left( i {\cal L}_2 \otimes \tilde {\cal N}_{n_0} \right)+ \texttt{span}\{ i {\bf 1}_l \otimes {\bf 1}_{n_0}\}.
 \ee
 \el
\vs The two cases $l >2$ and $l=2$, have to be treated
separately and this is done in the following two
subsections.

\subsection{Case $l>2$}
 \bl{Lemmal2} Assume
$l>2$. Then $\tilde {\cal N}_{n_0}$ (and therefore ${\cal N}_{n_0}$) is a Lie
algebra.  \el \bpr Assume without loss of generality $l=3$, since if
$l>3$ we can  assume in the following argument that all
elements which are not at the intersection of the first three rows
and columns, in the $l \times l$ matrices on the
left of the tensor products in $i{\cal L}_2\otimes \tilde {\cal N}_{n_0}$ of (\ref{s}),  are zero. Denote by $I_{j,k}$ and $R_{j,k}$, with $j<k
\in \{1,2,3\}$,  the matrix with all zeros except in the $(j,k)$-th
position which is occupied by $i$ or $1$, respectively
(correspondingly the $(k,j)$-th position is given). For any pair of
elements $N_1,N_2$ in $\tilde {\cal N}_{n_0}$ we calculate $[iI_{1,2}
\otimes N_1, iI_{1,3} \otimes N_2]$ which is in $[{\cal P}, {\cal
P}]$ because of (\ref{s}). Since elements in $\tilde {\cal N}_{n_0}$ are
skew-Hermitian and satisfy property (\ref{relat}), we have \be{asd}
Z:=[iI_{1,2} \otimes N_1, iI_{1,3} \otimes N_2]= \ee
$$
-\frac{1}{2}\left( \{ I_{1,2}, I_{1,3} \} \otimes [N_1, N_2] +
[I_{1,2}, I_{1,3}] \otimes \{ N_1, N_2\} \right)=
$$
$$
- \frac{1}{2} \left( i I_{2,3} \otimes [N_1,N_2]+ \alpha R_{2,3}
\otimes {\bf 1} \right),
$$
for some real $\alpha$.  By taking the Lie bracket with $R_{1,2}
\otimes {\bf 1}$ which is in ${\cal P}$, we obtain an element
in ${\cal P}$, which is given by \be{lastfr} [R_{1,2} \otimes {\bf
1}, Z]=-\frac{1}{2}\left( [R_{1,2}, I_{2,3}] \otimes [N_1, N_2]+
\alpha [R_{1,2}, R_{2,3}] \otimes {\bf 1}\right)= -\frac{1}{2}\left(
iI_{1,3} \otimes [N_1, N_2]+ \alpha R_{1,3} \otimes {\bf 1} \right)
\ee Since the last term in (\ref{lastfr}) is already in ${\cal P}$,
in order for $[R_{1,2} \otimes {\bf 1}, Z]$ to be in ${\cal P}$, we
must have $[N_1,N_2] \in \tilde {\cal N}_{n_0}$, that is, $\tilde {\cal
N}_{n_0}$ is closed under commutation.
 \epr

We have from Lemma \ref{Lemmal2} that ${\cal N}_{n_0}$  must be one of the Lie algebras  listed in Lemma \ref{fewLiealgebras}.\footnote{Recall that we are assuming that we have performed a change of coordinates so that  ${\cal N}_{n_0}$ has the canonical
form of Proposition \ref{classicnorm} and Lemma \ref{fewLiealgebras}.}  We can eliminate the first case which cannot be verified\footnote{Recall that we are assuming ${\cal P}$ non-Abelian.}  and the case where $n_0=1$ which would mean that ${\cal L}_S={\cal L}_1 +{\cal L}_2 + \texttt{span} \{ i {\bf 1}\}=u(l)=u(n_S)$ which we have excluded. In the case 2 of Lemma \ref{fewLiealgebras}, $\tilde {\cal N}_{n_0}=0$, so that  ${\cal P}={\cal L}_1\otimes {\bf 1}_{n_0} + \texttt{span} \{ i {\bf 1}_{n_S} \}  \otimes {\bf 1}_{n_0}$. Since $[{\cal L}_1, {\cal L}_1]={\cal L}_2$, from Corollary \ref{corollario1} (or by direct computation), we have $[{\cal P}, {\cal P}]={\cal L}_2 \otimes {\bf 1}_{n_0} \subseteq  {\cal K}$. Since ${\cal L}_S$ is reductive, from Lemma \ref{decoK}, we write
${\cal K}$ as ${\cal K}=({\cal L}_2 \otimes  {\bf 1}_{n_0}) + {\cal R}$ where ${\cal R}$ commutes with ${\cal P}$ and it is an ideal in ${\cal L}_S$. If we write a general element of ${\cal R}$ as $\sum_j R_j \otimes \sigma_j$, with $R_j \in u(l)$ and $\sigma_j$, $n_0 \times n_0$, Hermitian,  linearly independent matrices, we find that $[R_j, {\cal L}_1]=0$, which also (using $[{\cal L}_1, {\cal L}_1]={\cal L}_2$ and the Jacobi identity) implies $[R_j, {\cal L}_2]=0$, and therefore $[R_j, u(l)]=0$, which implies that $R_j$ is a multiple of the identity. Therefore ${\cal P}={\cal L}_1 \otimes {\bf 1}_{n_0} +  \texttt{span} \{ i {\bf 1}_{n_S} \}  \otimes {\bf 1}_{n_0}$ and ${\cal K}=({\cal L}_2 \otimes {\bf 1}_{n_0})  + ( {\bf 1}_l \otimes \tilde {\cal R})$ for some subalgebra $\tilde {\cal R}$ of $u(n_0)$. Consider now the vector space
\be{calVo}
{\cal V}:= \left(i {\cal L}_1 \otimes {\bf 1}_{n_0} \otimes {su(2)}\right) +
\left( {\cal L}_1 \otimes {\bf 1}_{n_0} \otimes {\bf 1}_2 \right) +
\ee
$$
\left( {\cal L}_2 \otimes {\bf 1}_{n_0} \otimes {\bf 1}_2 \right) +
 \left(i {\cal L}_2 \otimes {\bf 1}_{n_0} \otimes {su(2)}\right) +
 \left( {\bf 1}_l \otimes {\bf 1}_{n_0} \otimes su(2)\right).
$$
 By using formulas (\ref{newRiem}) and (\ref{antiRiem}) we can verify that $ad_{\cal L} {\cal V} \subseteq {\cal V}$.  Now consider initial states (recall that $ln_0=n_S$), $\rho_S=\frac{1}{ln_0}{\bf 1}_{ln_0}+i L_1 \otimes {\bf 1}_{n_0}$ for some $L_1 \in {\cal L}_1$, $L_1 \not=0$,  and arbitrary initial state for $A$, $\rho_A:=\frac{1}{2}{\bf 1} +i\sigma$, for some $\sigma \in su(2)$. The matrix
\be{irsra}
i \rho_S \otimes \rho_A = \frac{1}{2n_0l}\left( i {\bf 1}_{ln_0} \otimes {\bf 1}_2 - L_{1} \otimes {\bf 1}_{n_0} \otimes {\bf 1}_2 - {\bf 1}_{ln_0} \otimes \sigma -i L_{1} \otimes {\bf 1} \otimes \sigma \right)
\ee
belongs to ${\cal V} + \texttt{span} \{ i {\bf 1}_{ln_0} \otimes {\bf 1}_2 \}$, which is also invariant under $ad_{\cal L}$. Via direct computation, we get
$$
Tr_A  \left( {\cal V} + \texttt{span} (i {\bf 1}_{ln_0} \otimes {\bf 1}_2) \right) = {\cal L}_1 \otimes {\bf 1}_{n_0} + {\cal L}_2 \otimes {\bf 1}_{n_0} + \texttt{span} \{ i {\bf 1}_{n_S}\} \not= u(n_S),
$$
which contradicts Lemma \ref{Lemmabasic}.

The cases 3 and 4 of Lemma \ref{fewLiealgebras} are treated  with a similar technique. In
the case 3,  ${\cal P}$ is given by
\be{Pcase3}
{\cal P}=\left({\cal L}_1 \otimes {\bf 1}_{n_0} \right) + \left( {\cal L}_2 \otimes {\bf 1}_{r,s} \right) + \texttt{span} \{ i {\bf 1}_l \otimes {\bf 1}_{n_0} \}.
\ee
Calculating $[{\cal P}, {\cal P}]$ using (\ref{newRiem}) and Corollary \ref{corollario1}, we obtain
\be{calPP}
[{\cal P}, {\cal P}]={\cal L}_2 \otimes  {\bf 1}_{n_0} + {\cal L}_1 \otimes {\bf 1}_{r,s}.
\ee
The ideal ${\cal R}$ of Proposition \ref{decoK} has again the form ${\bf 1}_l \otimes {\tilde {\cal R}}$, where now $\tilde {\cal R}$ is a subalgebra of $u(n_0)$ which commutes with ${\bf 1}_{r,s}$ (and therefore spanned by  block diagonal matrices). If we consider the vector space
\be{calV12}
{\cal V}:=\left( u(l) \otimes {\bf 1}_{n_0} \otimes {\bf 1}_2 \right) +
\left( u(l) \otimes {\bf 1}_{r,s} \otimes {\bf 1}_2 \right) +
\left( i u(l) \otimes {\bf 1}_{n_0} \otimes su(2) \right) + \left( i u(l) \otimes {\bf 1}_{r,s} \otimes su(2) \right),
\ee
it is easy to check that this space is invariant under $ad_{\cal L}$. By considering  the initial condition
\be{jhas}
\rho_S \otimes \rho_A:=\left(\frac{1}{n_0 l}{\bf 1}_{n_0l}+iL \otimes {\bf 1}_{n_0} \right) \otimes \left(\frac{1}{2}{\bf 1}_2 + i \sigma \right),
\ee
for some $L  \in {i u(l)}$, $L \not=0$,  and any $\sigma \in su(2)$,  since
$i \rho_S \otimes \rho_A \in {\cal V}$, and $Tr_A({\cal V}) \not= u(n_S)=u(n_0 l)$ we find a contradiction with Lemma \ref{Lemmabasic}.
In the case 4, we must assume $n_0$ even and at least equal to 4, since if $n_0$ is equal to $2$,  using (\ref{basiscan2}) and Lemma \ref{espressioP},  ${\cal K} + {\cal P}=u(n_S)=u(2l)$ which we have excluded.
If $n_0 \geq 4$, the ideal ${\cal R}$ of Lemma \ref{decoK} has (using (\ref{intermsofPauli}) of Remark \ref{remar}) the form
${\cal R}={\bf 1}_l \otimes {\bf 1}_2 \times \tilde {\cal R}$ where $\tilde {\cal R}$ is a Lie subalgebra of $u(\frac{n_0}{2})$. We consider a vector space
\be{calV45}
{\cal V}:=\left(u(2l) \otimes {\bf 1}_{\frac{n_0}{2}} \otimes {\bf 1}_2\right)
+ i\left(u(2l) \otimes {\bf 1}_{\frac{n_0}{2}} \otimes su(2)\right),
\ee
which is invariant under $ad_{\cal L}$ and such that $Tr_A({\cal V})
\not= u(n_S)
=u(n_0 l)$. By taking an initial state $i \rho_S \otimes \rho_A \in {\cal V}$ we find again a contradiction with Lemma \ref{Lemmabasic}.



\subsection{Case $l=2$}
\label{l2conclu}

Recall Lemma \ref{espressioP} and
Proposition \ref{classicnorm}. Aside from the trivial case 1 of Proposition \ref{classicnorm},\footnote{This case would imply ${\cal P}$ Abelian which we have excluded.} the recursion described in this proposition ends with the case 2 or the case 3 for some appropriate $n$. If the recursion ends with case 2,  ${\cal P}$ is given by
\be{calP90}
{\cal P}:= +_{j=0}^{j_{max}} \texttt{span} \{ (i)^j(\sigma_x)^{\otimes j}
\otimes \{ \sigma_z, \sigma_y\} \otimes {\bf 1}_{n_j} \} \,
+ \, \texttt{span} \{ i {\bf 1}_{n_S} \},
\ee
where $n_j:=n_S 2^{-(j+1)}$.\footnote{With some abuse of notation we are using the notation $n_j$, here again as in Lemma \ref{Vandermonde}. However the meaning of $n_j$ for $j=1,...,j_{max}$ is different here than in that Lemma. In fact we are already in the situation where al the $n_j$'s of Lemma \ref{Vandermonde} are equal to $n_0$. In formula (\ref{calP90}) however $n_0$ coincides with the one previously defined.} The number $j_{max}$ is an integer number with ${j_{max}} \leq \log_2{n_S}-1$, which gives the number of iterations, i.e., how many times we return to step 2.   In order to see this,\footnote{We neglect here the factor $\frac{1}{2}$ in the definition of the Pauli matrices (\ref{Paulimat}) which has no effect on the vector spaces we are describing.} assume first that we reach step 2 and never come back. Then, in Lemma \ref{espressioP}, we only have ${\cal L}_1 \otimes {\bf 1}_{n_0}$, and $j_{max}=0$ and
the only linearly independent matrices to be included in a basis of  ${\cal P}$ are (beside the $i {\bf 1}_{n_S}$)
 $\sigma_y \otimes {\bf 1}_{\frac{n_S}{2}}$ and  $\sigma_z \otimes {\bf 1}_{\frac{n_S}{2}}$. However, if $\tilde {\cal N}_{n_0} \not=0$, we  move on to step 3 and have to add the matrix
$i \sigma_x \otimes i{\bf 1}_{\frac{n_S}{4}, \frac{n_S}{4}} =i \sigma_x \otimes \sigma_z \otimes {\bf 1}_{\frac{n_S}{4}}$ and, since we are supposed to go back to 2,  the matrix
\be{polka}
i \sigma_x \otimes \begin{pmatrix}{\bf 0} & {\bf 1}_{\frac{n_S}{4}} \cr -{\bf 1}_{\frac{n_S}{4}} & {\bf 0} \end{pmatrix}=i \sigma_x \otimes \sigma_y \otimes {\bf 1}_{\frac{n_S}{4}}.
\ee
Continuing this way we obtain the basis in (\ref{calP90}).

Anagously, in the case where the iteration ends with step 3,   we obtain for ${\cal P}$
\be{calP91}
{\cal P}:= +_{j=0}^{j_{max}} \texttt{span} \{ (i)^j \sigma_x^{\otimes j} \otimes \{\sigma_y, \sigma _z\} \otimes {\bf 1}_{n_j} \} \, \, +  \texttt{span} \left\{(i)^{j_{max}} \sigma_x^{j_{max} +1} \otimes {\bf 1}_{r,s} \right\}\, \, +
\texttt{span} \left\{{i {\bf 1}_{n_S}}\right\},
\ee
where $j_{max}$ is some nonnegative integer number with $j_{max} \leq \log_2 (n_S -2) -1$ and $r$ and $s$ are two nonnegative integer numbers with $r+s= n_S 2^{-(j_{max}+1)}$.

Consider the case (\ref{calP90}) first. If $n_{j_{max}} \geq 2$, we
have\footnote{This is trivially true even if $n_{j_{max}}=1$ but this case will be treated later.}
$$
[{\cal P}, {\cal P}] \subseteq u\left( \frac{n_S}{n_j} \right) \otimes {\bf 1}_{n_{j_{max}}}.
$$
Moreover, similarly to what described in the previous  subsection, the ideal  ${\cal R} \subseteq {\cal K}$ of (\ref{poi}), has the form ${\bf 1}_{\frac{n_S}{n_{j_{max}}}} \otimes \tilde {\cal R}$, for some subalgebra $\tilde {\cal R} \subseteq u(n_{j_{max}})$. The subspace
\be{subspac}
{\cal V}:= \left(u\left(\frac{n_S}{n_{j_{max}}}\right) \otimes {\bf 1}_{n_{j_{max}}} \otimes {\bf 1}_2 \right)
 +i\left(  u\left(\frac{n_S}{n_{j_{max}}}\right) \otimes {\bf 1}_{n_{j_{max}}} \otimes su(2) \right),
\ee
is invariant under $ad_{\cal L}$ and by taking an initial condition $\rho_S \otimes \rho_A$ of the form
\be{formainitcond}
\rho_S \otimes \rho_A=\left(\frac{1}{2}{\bf 1}_{n_S}+ iL \otimes {\bf 1}_{n_{j_{max}}} \right)
\otimes \left( \frac{1}{2} {\bf 1}_2 +i \sigma \right),
\ee
with $L$ a nonzero matrix in $su(\frac{n_S}{n_{j_{max}}})$ and $\sigma$ any matrix in $su(2)$, we find a contradiction with Lemma \ref{Lemmabasic}. Therefore $n_{j_{max}}$ must be $1$ in this case.   The same thing can be proved in  the  case (\ref{calP91}). If ${n_{j_{max}}} \geq 2$, then ${\cal P}$ and $[{\cal P}, {\cal P}]$ are subspaces of
$ u\left(\frac{n_S}{n_{j_{max}}} \right) \otimes \{\text{span} \{ {\bf 1}_{n_{j_{max}}}, {\bf 1}_{r,s}\} \} $, and the ideal  ${\cal R}$ of ${\cal K}$ defined in (\ref{poi}) has the form ${\bf 1}_{\frac{n_S}{n_{j_{max}}}} \otimes {\tilde {\cal R}}$, where now $\tilde {\cal R}$ has to commute with ${\bf 1}_{r,s}$. The space
\be{llo34}
{\cal V}:= \left( u \left( \frac{n_S}{n_{j_{max}}} \right) \otimes \{ {\bf 1}_{n_{j_{max}}}, {\bf 1}_{r,s} \} \otimes {\bf 1}_2 \right)  + i \left( u \left( \frac{n_S}{n_{j_{max}}} \right) \otimes \{ {\bf 1}_{n_{j_{max}}}, {\bf 1}_{r,s} \} \otimes su(2) \right),
\ee
is invariant under $ad_{\cal L}$ and, once again,  we find a contradiction with Lemma \ref{Lemmabasic}.

In conclusion, we have  to study only the cases (\ref{calP90}) and (\ref{calP91}) only for $n_{j_{max}}=1$, which is  $j_{max}=\log_2 n_S -1:=p$, assumed  integer. The dimension $n_S$ is equal to $2^{p+1}$, for some integer $p \geq 0$.   In the case  (\ref{calP90}), we have
\be{calP90Spec}
{\cal P}= +_{j=0}^{ p} \texttt{span} \{ (i)^j(\sigma_x)^{\otimes j} \otimes \{ \sigma_z, \sigma_y\} \otimes {\bf 1}_{n_j} \} \, + \texttt{span}\{i {\bf 1}_{n_S} \},
\ee
and, in the case  (\ref{calP91}),
\be{calP91bis}
{\cal P}= +_{j=0}^{p} \texttt{span} \{ (i)^j \sigma_x^{\otimes j} \otimes \{\sigma_y, \sigma _z\} \otimes {\bf 1}_{n_j} \} \, \, + \texttt{span} (i)^{p} \sigma_x^{j_{max} +1} \, \, \texttt{span}\{i {\bf 1}_{n_S} \}. \ee

We have therefore reduced the problem to the case where $n_S$ is equal to $n_S=2^{p+1}$, and $p+1$ is the number of factors in the tensor products of $2 \times 2$ matrices which span ${\cal L}_S$. Recall that we  denote by $\tilde {\cal P}$ the subspace of ${\cal P}$ of matrices with zero trace.
To be more explicit in the case (\ref{calP90Spec}), we have that $\tilde {\cal P}$ is the
span of the following matrices
\begin{eqnarray} \label{olp09}
&&\{\sigma_y, \sigma_z\} \otimes {\bf 1},   \\
&&i\sigma_x \otimes \{\sigma_y, \sigma_z\} \otimes {\bf 1},  \nonumber \\
&&\sigma_x \otimes \sigma_x \otimes \{\sigma_y, \sigma_z\} \otimes {\bf 1},  \nonumber \\
&&\vdots \nonumber \\
&&(i)^{p+1}\sigma_x \otimes \sigma_x \cdots \sigma_x \otimes \sigma_x \otimes \{\sigma_y, \sigma_z\}, \nonumber
\end{eqnarray}
while, in the case (\ref{calP91bis}), we have that $\tilde {\cal P}$ is the span of the following matrices
\begin{eqnarray} \label{olp10}
&&\{\sigma_y, \sigma_z\} \otimes {\bf 1},   \\
&&i\sigma_x \otimes \{\sigma_y, \sigma_z\} \otimes {\bf 1},  \nonumber \\
&&\sigma_x \otimes \sigma_x \otimes \{\sigma_y, \sigma_z\} \otimes {\bf 1}, \nonumber \\
&&\vdots \nonumber \\
&&(i)^{p+1}\sigma_x \otimes \sigma_x \cdots \sigma_x \otimes \sigma_x \otimes \{\sigma_y, \sigma_z, \sigma_x\}.  \nonumber
\end{eqnarray}

\vs

The proof can be carried out by considering separately  the cases $p=0, 1,2$ and then by induction for $p>2$. The case $p=2$ is quite long and it is postponed to Appendix B. The other cases are treated below.

\subsubsection{$p=0$ and $p=1$} If $p=0$, then both in the case (\ref{olp09}) and in the case (\ref{olp10}) ${\cal L}_S:={\cal P} + {\cal K}= u(n_S)$.\footnote{In this case, $n_0=1$ and $n_S=2$, and in the case (\ref{olp09}), $\tilde {\cal P}=\texttt{span} \{ \sigma_y, \sigma_z \}$, while in the case (\ref{olp10})  $\tilde {\cal P}=\texttt{span} \{ \sigma_x, \sigma_y, \sigma_z \}$. By using $[{\tilde {\cal P}}, {\tilde {\cal P}}] \subseteq {\cal K}$, we obtain that ${\cal L}_S=u(2)$.}  Therefore the condition we want to prove is automatically satisfied. If $p=1$, then calculating  $[{\cal P}, {\cal P}]= {\cal K}$, we find that in the case (\ref{olp10}) ${\cal L}_S=u(n_S)$ and therefore the theorem is automatically satisfied. In the case (\ref{olp09})
\be{calP0008}
{\cal P}:=\texttt{span} \left\{\{\sigma_y, \sigma_z\} \otimes {\bf 1}_2 \right\} + \left\{ \texttt{span} \{i \{\sigma_x \} \otimes \{\sigma_y , \sigma_z\}\}\right\} + \texttt{span} \{ i {\bf 1}_4 \}
\ee
\be{calP0009}
{\cal K}=[{\cal P}, {\cal P}]=\texttt{span}  \left\{i\{\sigma_y, \sigma_z\} \otimes \{\sigma_y, \sigma_z\} \right\} +
 \texttt{span} \{\sigma_x \otimes {\bf 1}_2 \ \} +  \texttt{span} \left\{{\bf 1}_2 \otimes \sigma_x \right\},
\ee
which is $10$-dimensional. Therefore the condition ${\cal L}_S=u(n_S)$ is not verified. We want to show that indirect controllability cannot be verified in this case. Once again consider the definition of  $\tilde {\cal P}$,  the subspace of ${\cal P}$ spanned by matrices with zero trace. Moreover define
\be{LS1LS2}
{\cal L}_{S,1}^\perp :=\texttt{span} \{ i \sigma_x \otimes \sigma_x \}, \qquad {\cal L}_{S,2}^\perp := \texttt{span} \{{\bf 1} \otimes \sigma_z, \, {\bf 1} \otimes \sigma_y, \, i \sigma_z \otimes \sigma_x, i \sigma_y \otimes \sigma_x\}.
\ee
By using (\ref{commurel}), (\ref{anticommurel}),\footnote{Along with
(\ref{tenscom}), (\ref{tensantcom})}  it is straightforward
to verify the following commutation relations,

\begin{eqnarray}
[\tilde {\cal P}, \tilde {\cal P}]& = &{\cal K}, \label{CR1JJJ} \cr
[\tilde {\cal P}, {\cal K}] & = & \tilde {\cal P}, \nonumber   \cr
[\tilde {\cal P}, {\cal L}_{S,1}^\perp]&=&{\cal L}_{S,2}^\perp, \nonumber  \cr
[\tilde {\cal P}, {\cal L}^\perp_{S,2}]&=&{\cal L}_{S,1}^\perp, \nonumber  \cr
[{\cal K}, {\cal K}]&=&{\cal K}, \nonumber   \cr
[{\cal K}, {\cal L}_{S,1}^\perp]&=&0, \nonumber   \cr
[{\cal K}, {\cal L}_{S,2}^\perp]&=&{\cal L}_{S,2}^\perp, \nonumber   \cr
[{\cal L}_{S,1}^\perp, {\cal L}_{S,1}^\perp]&=&0, \nonumber  \cr
[{\cal L}^\perp_{S,1}, {\cal L}_{S,2}^\perp]&=& \tilde {\cal P}, \nonumber   \cr
[{\cal L}_{S,2}^\perp, {\cal L}_{S,2}^\perp]&=&{\cal K},  \nonumber
\end{eqnarray}

\noindent and the anti-commutation relations
\begin{eqnarray}{}
i \{ \tilde {\cal P}, \tilde {\cal P} \}& = &  \texttt{span} \{ i\bf 1\}, \label{UI8} \cr
i \{ \tilde {\cal P}, {\cal K} \}& = &  {\cal L}_{S,2}, \nonumber  \cr
i \{ \tilde {\cal P}, {\cal L}_{S,1}^\perp  \}& = &  0, \nonumber  \cr
i \{ \tilde {\cal P}, {\cal L}_{S,2}^\perp  \}& = &  {\cal K}, \nonumber  \cr
i \{ {\cal K}, {\cal K} \}& = &  {\cal L}_{S,1}^\perp + \texttt{span} \{ i {\bf 1} \},  \nonumber  \cr
i \{ {\cal K}, {\cal L}^\perp_{S,1} \}& = &  {\cal K}, \nonumber  \cr
i \{ {\cal K}, {\cal L}^\perp_{S,2} \}& = &  \tilde {\cal P}, \nonumber  \cr
i \{ {\cal L}^\perp_{S,1}, {\cal L}^\perp_{S,1} \}& = &  \texttt{span} \{ {i \bf  1} \},  \nonumber  \cr
i \{ {\cal L}_{S,1}^\perp , {\cal L}^\perp_{S,2} \}& = &  0 \nonumber  \cr
i \{ {\cal L}_{S,2}^\perp , {\cal L}^\perp_{S,2} \}& = &  \texttt{span}  \{ i \bf 1 \}.  \nonumber
\end{eqnarray}

Consider now the vector space
\be{calVnuovo}
\bar {\cal V}:= {\cal L} + \left\{ {\cal K} \otimes (i \, \texttt{span}\{\sigma_x, \sigma_y,\sigma_z\}) \right\} + \left\{ {\cal P} \otimes {\bf 1}_2 \right\}
\ee
$$
 + \left\{{\cal L}_{S,2}^\perp \otimes  \left(i \, \texttt{span}\{\sigma_x, \sigma_y,\sigma_z\} \right) \right\} + \{ {\cal L}_{S,1}^\perp \otimes {\bf 1}_2 \}.
$$
From the fact that ${\cal L}$ is spanned by matrices of the form $K \otimes {\bf 1}$ with $K \in {\cal K}$ and $iP \otimes \sigma$, with $P \in {\cal P}$ and $\sigma$ any Pauli matrix, using the above commutation and anti-commutation relations, we verify that $\bar {\cal V}$
is invariant under ${\cal L}$, i.e., $[{\cal L}, \bar {\cal V}] \subseteq \bar {\cal V}$.

\noindent Consider now Lemma \ref{Lemmabasic} and pick initial
conditions $\rho_S$ and $\rho_A$ of the form $\rho_S=\frac{1}{4}{\bf 1} +K$, for a $K \in i{\cal K}$, $K \not=0$,  and $\rho_A=\frac{1}{2}{\bf 1} +\sigma$, with $\sigma \in i su(2)$. With this choice $i\rho_S \otimes \rho_A \in \bar {\cal V}$, and from invariance  ${\cal V}$ of Lemma \ref{Lemmabasic}, is such that ${\cal V} \subseteq \bar {\cal V}$. Since $Tr_A (\bar {\cal V}) \not= u(4)$, the necessary condition of Lemma \ref{Lemmabasic} is not satisfied, and therefore indirect controllability cannot be verified.

\subsubsection{$p=2$}

See Appendix B.

\subsubsection{$p>2$}
\label{plarge2}
${\cal P}$ in (\ref{olp09}) is a subspace of ${\cal P}$ in (\ref{olp10}), and a straightforward computation shows that, for the case (\ref{olp10}), $[{\cal P}, {\cal P}]={\cal K}$ (namely the ideal ${\cal R}$ of Lemma \ref{decoK} is $\{ { 0}\}$\footnote{This can be seen in both cases (\ref{olp09}) and (\ref{olp10}) imposing the fact that ${\cal R}$ commutes with ${\cal P}$ as from Lemma \ref{decoK}.} and ${\cal L}_S \not= u(8)$.  Therefore, it is enough to prove that indirect controllability cannot be verified in the case (\ref{olp10}).

Consider first the slightly more general case $p \geq 2 $.
To simplify the notations, we make a change of coordinates local on each one of the first $p$ positions so as to change the span of $\sigma_x$ into the span of $\sigma_z$ and viceversa and leave the span of $\sigma_y$ unchanged. We denote by  ${\cal P}_n$, $\tilde P$ for the case of $n:=p+1$ positions and ${\cal K}_n$, ${\cal K}$ in that case. By defining $Y:=\texttt{span} \{i \sigma_x, i \sigma_y\}$, $Z:=\texttt{span} \{ i\sigma_z \}$,
$\sigma:=\texttt{span}  \{i\sigma_x,i\sigma_y, i\sigma_z\}$, we have in particular\footnote{See formulas (\ref{recursrelat1}) and (\ref{recursrelat2}) below for a recursive expression  of ${\cal P}_n$ and ${\cal K}_n$.}
\be{calP3}
i\tilde {\cal P}:=i{\cal P}_3:= Y \otimes {\bf 1}\otimes {\bf 1} + Z \otimes Y \otimes {\bf 1} + Z \otimes Z \otimes \sigma,
\ee
\be{calK3}
i{\cal K}:=i{\cal K}_3=i [{\cal P}_3, {\cal P}_3]= {\bf 1} \otimes Z \otimes {\bf 1}+
Z \otimes {\bf 1} \otimes {\bf 1}+ Y \otimes Y \otimes {\bf 1}+ {\bf 1} \otimes 1 \otimes \sigma+ {\bf 1} \otimes Y \otimes \sigma+ Y \otimes Z \otimes \sigma,
\ee
so that ${\cal L}_S$,  in the case $p=2$,  can be taken equal to ${\cal L}_S={\cal K}_3 + {\cal P}_3 +  \texttt{span} \{i  {\bf 1}_8 \}$. Define the subspace of $su(8)$

\be{B3}
{\cal{B}}_{{3}}=i Y \otimes Z\otimes {\bf{1}} +   i {\bf{1}} \otimes Y\otimes {\bf{1}} +   Y \otimes {\bf{1}}\otimes su(2) +  Y\otimes Y \otimes su(2) +
\ee
$$
 {\bf{1}} \otimes Z\otimes  su(2)  +  Z\otimes Y \otimes su(2) +  Z\otimes {\bf{1}}\otimes su(2) +
i Z\otimes Z\otimes {\bf{1}}
$$
We have that
\[
su(8)= {\cal{K}}_{{3}} + {\cal{P}}_{{3}}+ {\cal{B}}_{{3}}.
\]
 We can verify the following commutation and anti-commutation relations:
\begin{itemize}

\item[(B1)]  $ i\left\{ {\cal{K}}_{3}, {\cal{P}}_{3} \right\}  = {\cal{B}}_{3}$

\item[(B2)]  $ i \left\{ {\cal{B}}_{3}, {\cal{P}}_{3} \right\}  = {\cal{K}}_{3}$

\item[(B3)]  $ \left[ {\cal{K}}_{3}, {\cal{B}}_{3} \right]  = {\cal{B}}_{3}$

\item[(B4)]  $ \left[ {\cal{P}}_{3}, {\cal{B}}_{3} \right]  = {\cal{B}}_{3}$

\item[(B5)] $i\left\{ {\cal P}_3, {\cal P}_3 \right\} = \texttt{span}\{ i {\bf 1}\}.$
\end{itemize}

\vs

From (\ref{olp10}) (and after the local change of coordinates defined above) it is straightforward to verify the following recursive relations.
\be{recursrelat1}
{\cal{P}}_{n+1}= Z\otimes {\cal{P}}_n +Y \otimes {\bf{1}}_2
\ee
\be{recursrelat2}
{ \cal{K}}_{n+1}= {\bf{1}}_2 \otimes {\cal{K}}_n+Y \otimes  {\cal{P}}_n + Z\otimes {\bf{1}}_2.
\ee
Using (B5) and (\ref{recursrelat1}) above,  by induction on $n$, we find, for every $n$,
\be{oolp45}
i \{ {\cal P}_n, {\cal P}_n \} =\texttt{span} \{ i {\bf 1} \}.
\ee
\bl{induzione1}
For any $n \geq 4$ there exist disjoint subspaces ${\cal{B}}_n$ and ${\cal{C}}_n$ such that
\begin{itemize}

\item[(A1)]  $ i \left\{ {\cal{K}}_n, {\cal{P}}_{n} \right\}  = {\cal{B}}_{n}$

\item[(A2)]  $ i \left\{  {\cal{P}}_n, {\cal{B}}_n \right\}  = {\cal{K}}_n$

\item[(A3)]  $ \left[ {\cal{K}}_n, {\cal{B}}_n \right]  = {\cal{B}}_n$

\item[(A4)]  $ \left[ {\cal{K}}_n, {\cal{C}}_n \right]  = {\cal{C}}_n$

\item[(A5)]  $ \left[ {\cal{P}}_n, {\cal{B}}_n \right]  = {\cal{C}}_n$

\item[(A6)]  $ \left[ {\cal{P}}_n, {\cal{C}}_n \right]  = {\cal{B}}_n$.
\end{itemize}
\el
\bpr We use  induction on $n$.  We first verify that (A1)-(A6) are satisfied for $n=4$. This can be done  using (\ref{recursrelat1}),  (\ref{recursrelat2}) and (B1)-(B5) and defining \be{calB4}
 i\left\{ {\cal{K}}_{4}, {\cal{P}}_{4} \right\}  =  Z\otimes {\cal{B}}_{3} +  Y\otimes  {\cal{K}}_{3} +   {\bf{1}} \otimes {\cal{P}}_{3} :=
 {\cal{B}}_4,
\ee
 and
\be{calC4}
 \left[ {\cal{P}}_{4}, {\cal B}_4 \right]  =  {\bf{1}} \otimes {\cal{B}}_{3} +  Y\otimes {\cal{B}}_{3}  + Z\otimes  {\cal{K}}_{3} := {\cal{C}}_4.
\ee
 Then we show that, if (A1)-(A6) hold for a certain $n$, they hold for $n+1$, which completes the proof by induction. In order to do that, define:
\be{Bnp1}
 {\cal{B}}_{n+1}:= i\left\{ {\cal{K}}_{{n}}, {\cal{P}}_{{n}} \right\} =  Z \otimes {\cal{B}}_n +  Y \otimes {\cal{K}}_n
 +  {\bf{1}} \otimes  {\cal{P}}_n,
\ee
and
\be{Cnp1}
 {\cal{C}}_{n+1} := \left[ {\cal{P}}_{{n+1}}, {\cal{B}}_{{n+1}} \right] =
{\bf{1}}\otimes  [{\cal{P}}_n, {\cal{B}}_n] +  Y \otimes {\cal{B}}_n + Z\otimes {\cal{K}}_n=
 {\bf{1}}\otimes  {\cal{C}}_n +  Y \otimes {\cal{B}}_n + Z\otimes {\cal{K}}_n.
\ee
  So both   (A1)  and (A5) are  automatically satisfied. Using  (\ref{oolp45}) we have:
  \[
 i \left\{{\cal{P}}_{{n+1}}, {\cal{B}}_{{n+1}} \right \}= i{\bf{1}} \otimes    \left\{{\cal{P}}_{{n}}, {\cal{B}}_{{n}} \right\} + Y\otimes {\cal{P}}_n  + i Z\otimes {\bf{1}}=   {\bf{1}}\otimes   {\cal{K}}_{{n}} + Y\otimes {\cal{P}}_n + iZ\otimes {\bf{1}}= {\cal{K}}_{n+1}.
  \]
  Therefore  (A2) holds. Now we verify (A3).
  \[
  \left[ {\cal{K}}_{{n+1}}, {\cal{B}}_{{n+1}} \right]=  Z\otimes  \left[ {\cal{K}}_{{n}}, {\cal{B}}_{{n}} \right]  + Y\otimes
  \left[ {\cal{K}}_n,{\cal{K}}_n\right]  +
  {\bf{1}} \otimes \left[{\cal{K}}_n,{\cal{P}}_n \right] + \]
  \[ +  i Y\otimes \left\{{\cal{P}}_{{n}}, {\cal{B}}_{{n}} \right\} + iZ\otimes \left \{ {\cal{K}}_{{n}}, {\cal{P}}_{{n}} \right\} +  {\bf{1}}\otimes \left[{\cal{P}}_n, {\cal{K}}_n\right] +  Y \otimes \left[{\cal{P}}_n, {\cal{P}}_n\right] +   Y \otimes  {\cal{K}}_n=
  \]
\[
= Z\otimes{\cal{B}}_n + Y \otimes  {\cal{K}}_n +  {\bf{1}} \otimes{\cal{P}}_n= {\cal{B}}_{n+1}.\]

\noindent Moreover we have
\[
 \left[ {\cal{K}}_{{n+1}}, {\cal{C}}_{{n+1}} \right]= {\bf{1}} \otimes{\cal{C}}_n +  Y \otimes {\cal{B}}_n + Z \otimes {\cal{K}}_n:= {\cal C}_{n+1}.
 \]
 Thus  (A4) holds. Next we verify  that (A6) holds:
 \[
 \left[ {\cal{P}}_{{n+1}}, {\cal{C}}_{{n+1}} \right]= Z\otimes \left[ {\cal{P}}_{{n}}, {\cal{C}}_{{n}} \right] +  Y \otimes
 i\left\{{\cal{P}}_{{n+1}}, {\cal{B}}_{{n+1}} \right\} +
  {\bf{1}} \otimes  \left[ {\cal{P}}_{{n}}, {\cal{K}}_{{n}} \right]
 + Z\otimes {\cal{B}}_n + Y\otimes {\cal{K}}_n=
 \]
\[= Z\otimes {\cal{B}}_n + Y \otimes {\cal{K}}_n +  {\bf{1}} \otimes  {\cal{P}}_{{n}}:=  {\cal{B}}_{n+1}.
\]
\epr

Given the above set-up the proof of the Theorem for the case $p>2$ is based
on the following observation.
\bl{induzione2}
Consider the Lie Algebra ${\cal L}_S={\cal K}_n + {\cal P}_n + \texttt{span}\{ i {\bf 1} \}$, and the  disjoint subspaces of $2^n \times 2^n$ matrices,
${\cal B}_n$ and ${\cal C}_n$, defined above,  so that, for every $n \geq 4$, the four disjoint subspaces  ${\cal{K}}_{{n}}$, $ {\cal{P}}_{{n}}$,
$ {\cal{B}}_{{n}}$, and ${\cal{C}}_{{n}}$ satisfy conditions (A1)-(A6) (besides  (\ref{oolp45})). Then the following space ${\cal V}$ is invariant
under ${\cal{L}}$,
\be{induzione3}
{\cal V}={\cal{K}}_{{n}}\otimes {\bf{1}}_2 +  i{\cal{K}}_{{n}}\otimes su(2) +  {\cal{P}}_{{n}}\otimes {\bf{1}}_2 +
i{\cal{P}}_{{n}} \otimes su(2)  +  i{\cal{B}}_{{n}} \otimes su(2) +  {\cal{C}}_{{n}}\otimes {\bf{1}}_2.
\ee
\el
\bpr
Using properties (A1)-(A6) and the definition (\ref{sommaYYY}), we verify that
$ \left[ {\cal{K}}_{{n}}\otimes {\bf{1}}, {\cal V} \right] \subseteq {\cal V}$,  $ \left[ i{\cal{P}}_{{n}} \otimes su(2), {\cal V} \right] \subseteq {\cal V}$,
and  $\left[ {\bf 1}_{2^n} \otimes su(2), {\cal V} \right] \subseteq {\cal V}$.
\epr

\vs

\noindent This Lemma allows us to conclude the proof for any $p >2$ ($n \geq 4$). Take an initial state $\rho_S={\bf{1}}+K$, with $K\in i{\cal{K}}_n$, and $\rho_A={\bf{1}}+\tilde \sigma$, with $\tilde \sigma \in isu(2)$.  Then:
\[
\rho_S \otimes \rho_A= {\bf{1}}_{2^n}\otimes {\bf{1}}_2+ {\bf{1}}_{2^{n}} \otimes \sigma + K\otimes {\bf{1}}_2+ K\otimes \sigma \in {\cal V} + \texttt{span} i \{{\bf{1}}\}_{2^{n+1}},
\]
where ${\cal V}$ is the subspace defined in equation(\ref{induzione3}). Since,  from Lemma \ref{induzione2},  ${\cal V} + \texttt{span} \{i\bf{1}_{2^{n+1}}\}$ is invariant under ${\cal L}$, we have that:
\[
\text{Tr}_A \left({\cal V} + \texttt{span} \{i{\bf{1}}_{2^{n+1}}\} \right)
= {\bf{1}} +  {\cal{K}}_n + {\cal{P}}_n + {\cal{C}}_n.
\]
This is not equal to $u(n_S)$ since ${\cal{B}}_n$ is missing, thus
contradicting the necessary condition of Lemma \ref{Lemmabasic}.
Therefore  the model is not indirectly controllable.

\vs

\section{Concluding Remarks}

It is possible to have full unitary control on a target system by controlling it indirectly via  an auxiliary system,  without having full controllability on the total system. The necessary and sufficient conditions for this to happen have been given in this paper. These conditions  are given in terms of the dynamical Lie algebra associated with the total system and the initial state of the auxiliary system. Further research is needed to design protocols for {\it constructive} indirect controllability, investigate indirect controllability in cases where there exists a {\it network} of quantum systems in between the auxiliary (fully controlled) system and the target system, and to  investigate {\it more general notions of indirect controllability}. These notions may be  given in terms not only of unitary maps but of more general completely positive maps. A weaker (not uniform)  notion of indirect controllability might also be useful in experiments where the initial state of $A$ and the transformation on the total system $S+A$ can be made dependent of the initial state of the system $S$. It is also important to investigate to what extent the introduction of an auxiliary control system can help in decoupling the target system from the environment in open quantum system control. We believe that the results and the framework developed here will be useful for the treatment of these problems as well.

\section*{Acknoledgement}
D. D'Alessandro and R. Romano research is partially supported by NSF under
 Grant ECCS0824085 and partially by ARO MURI under Grant W911NF-11-1-0268.

\label{ConRem}

\section*{Appendix A: Some proofs of the results in Section \ref{auxi}}

\subsection*{Proof of Lemma \ref{Vandermonde}}
\bpr All matrices in ${\cal A}$ can be
simultaneously diagonalized via a change of coordinates. So we can
assume that all matrices in ${\cal A}$ are diagonal. Consider a
basis of ${\cal A}$, ${\cal B}_S:=\{ A_1,\ldots, A_l\}$. Take any
element $A_j$ in the basis ${\cal B}_S$. We have that $A_j$ can be written as
\be{AJ6}
A_j:=\sum_k i \lambda_{j,k} \Pi_k^j,
\ee
where  $\Pi_k^j$  are diagonal projections and $\lambda_{j,k}$ are  all distinct eigenvalues. Since from (\ref{anticommurel2}) and the fact that ${\cal A}$ is maximal, we have $i \{{\cal A}, {\cal A} \} \subseteq {\cal A}$, it follows that if $i^{m-1} A_j^m \in {\cal A}$, $i^m A_j^{m+1}$ is also in ${\cal A}$, since $i\{A_j, i^{m-1} A_j^m\} \in {\cal A}$. Therefore $i \sum_k \lambda_{j,k}^m   \Pi_k^j \in {\cal A}$ for every $m \geq 0$ (since ${\cal A}$ also contains multiples of the identity). A Vandermonde determinant argument, using the fact that the $\lambda_{j,k}$'s are all different, shows that the diagonal projections  $\Pi_k^j$ also belong to ${\cal A}$. Repeating this argument for all $A_j$'s we find a set of diagonal projections which (multiplied by $i$) span ${\cal A}$. In this set, choose a maximal linearly independent set $\{i \Pi_1,\ldots,i \Pi_l \}$. Starting from the set $\{i \Pi_1,\ldots,i \Pi_l \}$ it is possible to construct another spanning set for ${\cal A}$, of the form $\{i \tilde \Pi_1,\ldots,i \tilde \Pi_s \}$, with $s \geq l$, and $\tilde \Pi_j$ all diagonal projections,  with the property that \be{zerointersection}\tilde \Pi_j \tilde \Pi_k= \delta_{j,k} \tilde \Pi_j.
   \ee
This is done recursively starting from the set  $\{ \Pi_1,\ldots, \Pi_l \}$. Given two projections, say $\Pi_1$ and $\Pi_2$, we can replace them in the set with three projections $\Pi_1 \Pi_2$, $\Pi_1-\Pi_1 \Pi_2$ and $\Pi_2 -\Pi_1 \Pi_2$, which still span the subspace spanned by $\Pi_1$ and $\Pi_2$, are such that when multiplied by $i$ belong to ${\cal A}$, because of the property $i\{{\cal A}, {\cal A} \} \subseteq {\cal A}$, and have the property that the product of any pair of them give zero. Repeating this process recursively, we obtain a spanning set,  $\{i \tilde \Pi_1,\ldots,i \tilde \Pi_s \}$, for ${\cal A}$, with all the products between different projections equal to zero. A basis is obtained choosing a minimal spanning set in this set. The basis (\ref{favbasis}) is obtained after a change of coordinates which groups together the $1$'s in the same matrix.
\epr

\vs

\subsection*{Proof of Lemma \ref{decoK}}

\bpr First of all write ${\cal L}_S$ as ${\cal L}_S=[{\cal L}_S, {\cal L}_S] + {\cal A}_S$, where $[{\cal L}_S, {\cal L}_S]$ is the semisimple part of ${\cal L}_S$  and ${\cal A}_S$ the Abelian part. Observe that if $Y \in  [{\cal L}_S, {\cal L}_S]$ and $\langle Y, Y \rangle_K=0$ then $Y=0$ since the restriction of the Killing form on  $[{\cal L}_S, {\cal L}_S]$ is equal to the Killing form on this semisimple Lie algebra which is (negative) definite.

Now, given a basis in $[{\cal P}, {\cal P}]$ complete it in ${\cal K}$ with matrices $\{R_1,\ldots,R_r\}$ which are orthogonal to $[{\cal P}, {\cal P}]$ with respect to the Killing form and set ${\cal R}:=\texttt{span}\{ R_1, \ldots, R_r\}$. Let $R \in {\cal R}$ and $P_1,P_2 \in {\cal P}$. We have

\be{ideal1} \langle [R,P_1], P_2 \rangle_K= \langle [P_1, P_2], R
\rangle_K=0, \ee which says that $[R,P_1] \in {\cal P}^\perp $.\footnote{Orthogonality is meant with respect to the Killing form.} However $[R,P_1] \in [{\cal R}, {\cal P}] \subseteq [{\cal K},
{\cal P}] \subseteq {\cal P}$. Therefore $[{R},{P}_1] \in {\cal P} \cap {\cal P} ^\perp$. Since $[{R},{P}_1] \in [{\cal L}_S, {\cal L}_S]$ and the Killing form is negative definite in $[{\cal L}_S, {\cal L}_S]$, necessarily $[R,P_1]=0$. Thus
${\cal R}$ commutes with ${\cal P}$.

Now we show that ${\cal R}$ is also an ideal in ${\cal K}$.
Let  $R$ be an arbitrary  element in ${\cal R}$,
and $K$ an arbitrary element in ${\cal K}$ and
$P_1$ and $P_2$ arbitrary  elements in ${\cal P}$. Using the invariance property (\ref{Jacobiplus}) and the Jacobi
identity for Lie algebras we have \be{ideal2} \langle [K, R], [P_1,
P_2] \rangle_K= \langle [[P_1, P_2], K], R \rangle_K= - \langle
[[P_2,K], P_1], R \rangle_K - \langle [[K, P_1], P_2], R
\rangle_K=0, \ee where the last equality follows from $[{\cal K},
{\cal P}] \subseteq {\cal P}$ and the fact that  ${\cal R}$ is orthogonal with respect to the Killing form to $[{\cal P}, {\cal P}]$. Therefore $[K, R]$ not only belongs
to ${\cal K}$ but it is also
orthogonal  to $[{\cal P}, {\cal P}]$. Now write $[K,R]$ as $[K,R]=Y+\tilde R$, with $Y \in [{\cal P}, {\cal P}]$ and $\tilde R \in {\cal R}$. Since $\tilde R \in [{\cal P}, {\cal P}]^\perp$ and
 $[K,R] \in [{\cal P}, {\cal P}]^\perp$, $Y \in [{\cal P}, {\cal P}]^\perp$ as well. Therefore we have $Y \in [{\cal L}_S, {\cal L}_S]$ and $Y \in [{\cal P}, {\cal P}] \cap  [{\cal P}, {\cal P}]^\perp$, which as before implies $Y=0$. Therefore $[K,R]=\tilde R \in {\cal R}$ and ${\cal R}$ is an ideal in ${\cal K}$
 \epr

\section*{Appendix B: Case $p=2$ in subsection \ref{l2conclu}}
\label{peq2ff}
The same considerations done at the beginning of subsection \ref{plarge2} hold for the case $p=2$ to argue that it is enough to prove that indirect controllability is not verified in the case (\ref{olp10}). We also use the notation and the change of coordinates described  at the beginning of subsection \ref{plarge2}.

Observe that the dynamical Lie algebra ${\cal L}$ admits a Cartan decomposition,\footnote{Recall  that we are including now the auxiliary system $A$ in the analysis. The matrices in ${\cal L}$ are $16 \times 16$.}
\be{Cartandec}
{\cal L}:=\hat {\cal K} + {\hat  {\cal P}},
\ee
with $\hat {\cal K}:=\left({\cal K}_3 \otimes {\bf 1} \right) + \left( {\bf 1}_8  \otimes su(2) \right)$ and
$\hat  {\cal P}=i{\cal P}_3 \otimes su(2)$. Using such a Cartan decomposition, a general transformation $U$ in $e^{\cal L}$ can be parametrized as\footnote{See the discussion at the end of subsection \ref{specres}.} $U=T_2 \otimes V_2 e^{\tilde A} T_1 \otimes V_1$, where $T_1$  and $T_2$ are general unitary transformations in $e^{{\cal K}_3}$, $V_1$ and $V_2$ are general matrices in $SU(2)$ and $\tilde A$ is a matrix in a Cartan subalgebra $\cal A$ (maximal Abelian subalgebra) in $\hat {\cal P}$. A general unitary matrix $U$ in $e^{\cal L}$ gives a transformation on the state $\rho_S$ of the form,
\be{genU2}
\rho_S \rightarrow Tr_A(U \rho_S \otimes \rho_A U^\dagger)=Tr_A\left(T_2 \otimes V_2 e^{\tilde A} T_1 \otimes V_1 \rho_S \otimes \rho_A T_1^\dagger  \otimes V_1^\dagger  e^{-\tilde A} T_2^\dagger  \otimes V_2^\dagger \right)=
\ee
$$
=T_2Tr_A \left(e^{\tilde A} (T_1 \rho_S T_1^\dagger \otimes  \tilde \rho_A) e^{-\tilde A} \right) T^{\dagger}_2,
$$
with $\tilde \rho_A:=V_1 \rho_A V_1^\dagger$.
In this case, the Cartan subalgebra $\cal A$ is $3-$ dimensional. We define (cf. (\ref{Paulimat}))
\be{newPaulimat5}
\tsx:=-2i \sigma_x, \qquad \tsy:=2i \sigma_y, \qquad \tsz:=-2i \sigma_z,
\ee
and we take as a basis of ${\cal A}$,
$\{ i \tsz \otimes \tsz \otimes \tsx \otimes \tsx,
i \tsz \otimes \tsz \otimes \tsy \otimes \tsy,
i \tsz \otimes \tsz \otimes \tsz \otimes \tsz\}$, so that $\tilde A$ in (\ref{genU2}) is written as
\be{formaAA}
\tilde A:=i x \tsz \otimes \tsz \otimes \tsx \otimes \tsx +iy \tsz \otimes \tsz \otimes \tsy \otimes \tsy +
iz \tsz \otimes \tsz \otimes \tsz \otimes \tsz,
\ee
for real parameters $x,$ $y$ and $z$.
Moreover,  since $\tilde \rho_A$ is assumed pure according to what
proved in subsection \ref{purenec}, we can write $\tilde \rho_A$ as
\be{parametrizrhoA}
\tilde \rho_A:=\begin{pmatrix} \cos^2 (\theta) & -\frac{1}{2} \sin(2 \theta) e^{it} \cr
-\frac{1}{2} \sin(2 \theta) e^{-it} & \sin^2(\theta) \end{pmatrix},
\ee
for parameters $\theta$ and $t$ in $\RR$.
Formula (\ref{genU2}) describes the set of available transformations
on $\rho_S$. Each of these transformations  can be
seen as the cascade of three transformations:
 \begin{enumerate}
 \item A unitary transformation $\rho \rightarrow T_1 \rho T^\dagger_1$, with $T_1 \in e^{{\cal K}_3}$ and therefore depending on $21=\dim \,  {\cal K}_3$ parameters.

 \item A, not necessarily unitary, transformation $\rho \rightarrow Tr_A(e^{\tilde A} \rho \otimes \tilde \rho_A e^{-\tilde A})$, which depends on $5$ parameters, i.e.,  $x,y,z$ in (\ref{formaAA}) and $\theta$ and $t$ in (\ref{parametrizrhoA}).

\item Another unitary transformation $\rho \rightarrow T_2 \rho T^\dagger_2$, with $T_2 \in e^{{\cal K}_3}$ and therefore depending on $21=\dim {\cal K}_3$ parameters.

\end{enumerate}
To prove the claim it is enough to show that there is a unitary similarity transformation $X_f$, $\rho_S \rightarrow X_f \rho_S X_f^\dagger$, which cannot be obtained as the cascade of the above three transformations, no matter what parameters are chosen in the various steps. We shall show that this is the case for the transformation $X_f=X_{1-2}$ which switches the first and second position in a tensor product of three $2 \times 2$ Hermitian matrices $\tilde \sigma_1,$ $\tilde \sigma_2$, $\tilde \sigma_3$, i.e.,
\be{asdf}
X_{1-2} \tilde \sigma_1 \otimes \tilde \sigma_2 \otimes \tilde \sigma_3 X_{1-2}^\dagger = \tilde \sigma_2 \otimes \tilde \sigma_1 \otimes \tilde \sigma_3, \, \, \forall  \tilde \sigma_1, \tilde \sigma_2, \tilde \sigma_3 \, \in iu(2)
\ee

\vs

Let us set-up few more definitions.

\vs

\noindent With spaces of $4 \times 4$ Hermitian matrices  \be{LandR}
{\bf L}:= \{{\bf 1} \otimes Z \} + \{ Z \otimes {\bf 1} \} + \{Y \otimes Y\}, \qquad {\texttt{and}}  \qquad {\bf R}:= \{ {\bf 1} \otimes Y \} + \{ Y \otimes Z \}, \ee
we can rewrite $i {\cal K}_3$ as
\be{iK3}
i{\cal K}_3:= \{{\bf L} \otimes {\bf 1}\} + \{ {\bf R} \otimes (isu(2))\}
+ \{ {\bf 1} \otimes (i su(2)) \}.
\ee
 Consider now a general matrix $\tilde \rho_S$ of the form
 $\tilde \rho_S=\frac{1}{8} {\bf 1}_8+S$, with $S \in i {\cal K}_3$. Such a matrix, can be written as
\be{rhosforproof}
\tilde \rho_S:= \frac{1}{8} {\bf 1}_8+ L \otimes {\bf 1}_2 + \sum_{j=x,y,z} R_j \otimes \tilde \sigma_{j} + \sum_{k=x,y,z} a_k {\bf 1}_4 \otimes \tilde \sigma_{k},
\ee
with $L \in {\bf L}$, $R_{x,y,z} \in {\bf R}$ and $a_{x,y,z}$ real numbers. For such a type of matrix, we
calculate explicitly $Tr_A(e^{\tilde A} \tilde \rho_S\otimes \tilde \rho_A  e^{-{\tilde A}})$. Using the definition where $E_{j,k}$ is the $ 4 \times 4$ matrix with all zeros except for the entries $j$ and $k$ on the diagonal which are occupied by $1$, we obtain
\be{explicitYYY}
Tr_A(e^{\tilde A} \tilde \rho_S \otimes \tilde \rho_A e^{-\tilde A})=\frac{1}{8} {\bf 1}_8+
(\sin^2(y)-\sin^2(x))\cos(2 \theta) {\bf 1}_4 \otimes \tsz +
\ee
$$
\sin(2 \theta) \sin(2z) \cos(t) \sin(x-y) {\bf 1}_4 \otimes \tsx +
\sin(2 \theta) \sin(2z) \sin(t) \sin(x+y) {\bf 1}_4 \otimes \tsy+
$$
$$
L \otimes {\bf 1}_2 + (\sin^2(y)-\sin^2(x))\cos(2 \theta) L \otimes \tsz +
\sin(2 \theta) \sin(2z) \cos(t) \sin(x-y) L \otimes \tsx +
$$
$$
\sin(2 \theta) \sin(2z) \sin(t) \sin(x+y) L \otimes \tsy+
\frac{1}{2} \cos(2 \theta) \cos(2z)( \cos(2x)- \cos(2y)) R_z \otimes {\bf 1}_2 +
$$
$$
\cos(2z) R_z \otimes \tsz +
\frac{1}{2} \cos(2 \theta) \sin(2z) (\cos(2x)+ \cos(2y))\left(iE_{1,4}R_z E_{2,3}-iE_{2,3}R_z E_{1,4}\right) \otimes {\bf 1}_2+
$$
$$
\frac{1}{2}\sin(2z) \sin(2 \theta) \cos(t) \left(\sin(2x) - \sin(2y) \right) R_x \otimes {\bf 1}_2+
\cos(x+y) R_x \otimes \tsx+
$$
$$
-\frac{1}{2}\cos(2z) \sin(2 \theta) \cos(t)(\sin(2x)+\sin(2y))\left( iE_{1,4}  R_x E_{2,3} - i E_{2,3} R_x E_{1,4} \right) \otimes {\bf 1}_2+
$$
$$
\frac{1}{2} \sin(2 \theta) \sin(t) \sin(2z) \left( \sin(2x)+ \sin(2y) \right) R_y \otimes {\bf 1}_2+
$$
$$
\cos(x-y) R_y \otimes \tsy-
\frac{1}{2} \sin(2 \theta) \sin(t) \cos(2z) \left( \sin(2x)- \sin(2y) \right) \left( i E_{1,4} R_y E_{2,3}- i E_{2,3} R_y E_{1,4} \right) \otimes {\bf 1}_2+
$$
$$
a_z \left( \cos^2(x)- \sin^2(y) \right) {\bf 1}_4 \otimes \tsz -
a_z \sin(2 \theta) \cos(2z) \sin(t) \sin(y-x) \tsz \otimes \tsz \otimes \tsx +
$$
$$
a_z \sin(2 \theta) \cos(2z) \cos(t) \sin(x+y) \tsz \otimes \tsz \otimes \tsy+
a_x \cos(2z) \cos(x-y) {\bf 1}_4 \otimes \tsx -
$$
$$
\frac{a_x}{2}(\sin(2x) - \sin(2y)) \sin(2 \theta) \sin(t) \tsz \otimes \tsz \otimes \tsz -
a_x \sin(2z) \cos(x+y) \cos(2 \theta) \tsz \otimes \tsz \otimes \tsy +
$$
$$
a_y \cos(2z) \cos(x+y) {\bf 1}_4 \otimes \tsy -
\frac{a_y}{2}\cos(t) \sin(2 \theta)( \sin(2y)+ \sin(2x)) \tsz \otimes \tsz \otimes \tsz +
$$
$$
a_y \sin(2z) \cos(x-y) \cos(2 \theta) \tsz \otimes \tsz \otimes \tsx.
$$
Since $X_{1-2}$ is unitary, the transformation in (\ref{explicitYYY}) must be unitary and in particular it must leave multiples of the identity unchanged. This implies that we have in  (\ref{explicitYYY})
\be{CONDI1}
(\sin^2(y)-\sin^2(x))\cos(2 \theta)=0,
\ee
\be{CONDI2}
\sin(2 \theta) \sin(2z) \cos(t) \sin(x-y)=0,
\ee
\be{CONDI3}
\sin(2 \theta) \sin(2z) \sin(t) \sin(x+y)=0.
\ee
This also implies that if $\tilde \rho_S$ is of the form $\tilde \rho_S=\frac{1}{8} {\bf 1}_8+ L \otimes {\bf 1}_2$ with $L \in {\bf L}$, it is left unchanged by the transformation (\ref{explicitYYY}). From this, we try to obtain information on the form of $T_1$. Denote by $\rho_{ini}:=\rho_S-\frac{1}{8}{\bf 1}_8$ and by $\rho_{fin}=X_{1-2}{\rho_S} X_{1-2}^\dagger - \frac{1}{8}{\bf 1}_8$. Assume that $\rho_{ini} \in T^\dagger_1 {\bf L} \otimes {\bf 1}_2 T_1$. Since $T_{1} \in e^{{\cal K}_3}$, $\rho_{ini}$ belongs to $i {\cal K}_3$. Moreover, since $T_{2} \in e^{{\cal K}_3}$, $\rho_{fin} \in i {\cal K}_3$ as well. Since
$\rho_{fin}$ is obtained from $\rho_{ini}$ by switching the first and second position in the tensor products, $\rho_{ini}$ must belong to the subspace of $i{\cal K}_3$ which remains in $i{\cal K}_3$ once we permute the first two positions. This subspace is given by $\left\{{\bf L} \otimes {\bf 1}_2 \right\} + \left\{{\bf 1}_4 \otimes i su(2) \right\}$. This reasoning shows that $T_1$ is such that
\be{importantconditione}
T_1^\dagger {\bf L} \otimes {\bf 1} T_1 \subseteq \left\{{\bf L} \otimes {\bf 1}_2 \right\} + \left\{{\bf 1}_4 \otimes i su(2) \right\}.
\ee
Now we proceed to a parametrization of $T_1$
according to a Cartan decomposition of ${\cal K}_3$.  Let
${\cal K}_3:= {\cal D} + {\cal Q}$, with ${\cal D}:= \left\{ i{\bf L} \otimes {\bf 1}_2 \right\} + \left\{{\bf 1}_4 \otimes  su(2)\right\}$ and ${\cal Q}:= {\bf R} \otimes su(2)$, with (cf. (\ref{Riem}))
\be{hui33}
[{\cal D}, {\cal D}] \subseteq {\cal D}, \qquad [{\cal Q}, {\cal D}] \subseteq {\cal Q},
\qquad [{\cal Q}, {\cal Q}] \subseteq {\cal D}.
\ee
Choosing a basis of a Cartan subalgebra in ${\cal Q}$ given by $\{ i \tsx \otimes \tsz \otimes \tsx, i \tsy \otimes \tsz \otimes \tsy, i {\bf 1}_2 \otimes \tsx \otimes \tsz \}$, we write $T_1 \in e^{{\cal K}_3}$ as
\be{tt78}
T_1 := P_1 \otimes \tilde V_1 e^{B} P_2 \otimes \tilde V_{2},
\ee
with $P_1, P_2 \in e^{i {\bf L}}$, $\tilde V_1, \tilde V_2 \in SU(2)$, and
\be{BBBB}
B:= ia \tsx \otimes \tsz \otimes \tsx + ib \tsy \otimes \tsz \otimes \tsy + ic {\bf 1}_2 \otimes \tsx \otimes \tsz,
 \ee
for real parameters $a,b,$ and $c$.  With the structure of $T_1$ in (\ref{tt78}), condition (\ref{importantconditione}) implies that, for every $L \in {\bf L}$,
\be{anc}
e^{-B} L \otimes {\bf 1}_2 e^{B} = M \otimes {\bf 1}_2 + {\bf 1}_4 \otimes \tilde \sigma,
\ee
for some $M \in {\bf L}$ and $\tilde \sigma \in su(2)$. Imposing this
for a basis of ${\bf L}$, we find,  for
the parameters $a$, $b$, and $c$ in (\ref{BBBB}),
\be{constraintsss}
\sin(2a)=\sin(2b)=\sin(2c)=0.
\ee
This, by writing $e^{B}$ as
\be{neweB}
e^{B}= e^{ia \tsx \otimes \tsz \otimes \tsx} e^{ib \tsy \otimes \tsz \otimes \tsy} e^{ic {\bf 1}_2 \otimes \tsx \otimes \tsz},
\ee
implies that the first factor is equal to $\pm {\bf 1}_8$ or $\pm i \tsx \otimes \tsz \otimes \tsx$, the second factor is equal to $\pm {\bf 1}_8$ or $\pm i \tsy \otimes \tsz \otimes \tsy$ and the third factor is equal to $\pm {\bf 1}_8$ or  $\pm i {\bf 1}_2 \otimes \tsx \otimes \tsz$. In particular,  in every case $e^{B}$,  has the form of a 'local' transformation
\be{local8}
e^{B}=C_1 \otimes C_2 \otimes C_3,
\ee
with unitary $2 \times 2$ transformations $C_1$, $C_2$, $C_3$. This, combined with (\ref{tt78}),   shows that $T_1$ must be of the form
\be{formaT1}
T_1:= Q_1 \otimes \tilde V_1 C_3 \tilde V_2,
\ee
with $Q_1:=P_1 (C_1 \otimes C_2) P_2.$

Let us now apply the full cascade of the three transformations in points 1, 2, and 3 above,  which, by assumption, gives $X_{1-2}$, to
${\bf 1}_4 \otimes   (\tilde V_2^\dagger C_3^\dagger \tilde V_1^\dagger \tsz \tilde V_1 C_3 \tilde V_2)$. Application of the transformation $\rho \rightarrow T_1 \rho T_1^\dagger $ gives ${\bf 1} \otimes \tsz$. By applying (\ref{explicitYYY}) to ${\bf 1} \otimes \tsz$ with the conditions (\ref{CONDI1}), (\ref{CONDI2}), (\ref{CONDI3}), we have
\be{expliZZ}
Tr_A(e^{\tilde A } {\bf 1}_4 \otimes \tsz \otimes \tilde \rho_A e^{-{\tilde A}})=
\left( \cos^2(x)- \sin^2(y) \right) {\bf 1}_4 \otimes \tsz -
\ee
$$
\sin(2 \theta) \cos(2z) \sin(t) \sin(y-x) \tsz \otimes \tsz \otimes \tsx +
\sin(2 \theta) \cos(2z) \cos(t) \sin(x+y) \tsz \otimes \tsz \otimes \tsy.
$$
Since the result of this transformation must be
in $i{\cal K}_3$, because the third transformation $T_2 \in e^{ {\cal K}_3}$
and $X_{1-2} {\bf 1}_4 \otimes \tsz X_{1-2}^\dagger={\bf 1}_4 \otimes \tsz \in i{\cal K}_3 $, we must have
$
\sin(2 \theta) \cos(2z) \sin(t) \sin(y-x)=0,
$ and  $
\sin(2 \theta) \cos(2z) \cos(t) \sin(x+y)=0.
$
These imply in (\ref{expliZZ}), since the norm has to be
preserved (because the total transformation must be unitary),
\be{K1}
(\cos^2(x)- \sin^2(y))^2=1.
\ee
This condition along with (\ref{CONDI1}), (\ref{CONDI2}), (\ref{CONDI3}), gives the following simplification of (\ref{explicitYYY})
\be{simplifiedexplicitYYY}
Tr_A(e^{\tilde A} \tilde \rho_S \otimes \tilde \rho_A e^{-\tilde A})=\frac{1}{8} {\bf 1}_8+
L \otimes {\bf 1}_2 +
\cos(2z) R_z \otimes \tsz
\ee
$$\pm \cos(2 \theta) \sin(2z) \left(iE_{1,4}R_z E_{2,3}-iE_{2,3}R_z E_{1,4}\right) \otimes {\bf 1}_2\pm R_x \otimes \tsx
$$
$$ \pm R_y \otimes \tsy
\pm a_z  {\bf 1}_4 \otimes \tsz \pm a_x \cos(2z) {\bf 1}_4 \otimes \tsx
$$
$$\pm a_x \sin(2z)  \cos(2 \theta) \tsz \otimes \tsz \otimes \tsy \pm a_y \cos(2z) {\bf 1}_4 \otimes \tsy\pm a_y \sin(2z) \cos(2 \theta) \tsz \otimes \tsz \otimes \tsx.
$$
Now assume we start from ${\bf 1}_4 \otimes   (\tilde V_2^\dagger C_3^\dagger \tilde V_1^\dagger \tsx \tilde V_1 C_3 \tilde V_2)$. Application of the transformation $\rho \rightarrow T_1 \rho T_1^\dagger $ gives ${\bf 1} \otimes \tsx$. Using (\ref{simplifiedexplicitYYY}) with $\tilde \rho_S= \frac{1}{8}{\bf 1}_8+{\bf 1}_4 \otimes \tsx$ gives
\be{expliXX}
Tr_A(e^{\tilde A} \left( {\bf 1}_4 \otimes \tsx \otimes \tilde \rho_A \right) e^{-\tilde A})=\pm
\cos(2z)  {\bf 1}_4 \otimes \tsx \pm \sin(2z)  \cos(2 \theta) \tsz \otimes \tsz \otimes \tsy.
\ee
Imposing that this belongs to $i{\cal K}_3$, gives
\be{K3}
   \sin(2z)  \cos(2 \theta)     =0.
\ee
Moreover norm preservation gives $\cos^2(2z)=1$. Using this in (\ref{simplifiedexplicitYYY})
 we get
 \be{explicitSSSS}
Tr_A(e^{\tilde A} \tilde \rho_S \otimes \tilde \rho_A e^{-{\tilde A}})=\frac{1}{8} {\bf 1}_8+
L \otimes {\bf 1}_2 \pm
\ee
$$
R_z \otimes \tsz \pm R_x \otimes \tsx \pm R_y \otimes \tsy
\pm a_z {\bf 1}_4 \otimes \tsz \pm a_x  {\bf 1}_4 \otimes \tsx
\pm a_y  {\bf 1}_4 \otimes \tsy.
$$
Therefore $\tilde \rho_S \rightarrow Tr_A(e^{\tilde A} \tilde \rho_S \otimes \tilde \rho_A  e^{-{\tilde A}})$, does not modify $\tilde \rho_S$ except for possibly some changes in the sign of the coefficients. It follows that if $\tilde \rho_S =\frac{1}{8}{\bf 1}_8 +S$ with $S \in i{\cal K}_3$, the transformed also can be written as ${\bf 1}_8+ \tilde S$, with $\tilde S \in i {\cal K}_3$. It follows that if the initial $\rho_S$ has the property that $\rho_S -\frac{1}{8} {\bf 1}_8 \in i{\cal K}_3$, the final value of the density matrix has this property as well (since the similarity  transformations by $T_1$ and $T_2$ do not modify the property of a matrix to belong to $i{\cal K}_3$). However this is incompatible with the form of $X_{1-2}$ since the transformation $\rho \rightarrow X_{1-2} \rho X_{1-2}^\dagger$, does {\it not} leave $i{\cal K}_3$ invariant. This concludes the proof of this part of the Theorem.

\end{document}